\documentstyle [11pt,epsf] {article}

\begin{document}

\begin{center}
\vspace{2cm}
\LARGE
Gas Infall and  Stochastic Star Formation in Galaxies  
in the Local Universe                 
\\                                                     
\vspace{1cm} 
\large
Guinevere Kauffmann$^1$, Timothy M. Heckman$^2$, Gabriella De Lucia$^1$,\\                   
Jarle Brinchmann$^3$, St\'ephane Charlot$^4$, Christy Tremonti$^5$, Simon D.M. White$^1$  \\
Jon Brinkmann$^6$ \\
\vspace{0.5cm}
\small
\noindent
{\em $^1$Max-Planck Institut f\"{u}r Astrophysik, D-85748 Garching, Germany} \\
{\em $^2$Department of Physics and Astronomy, Johns Hopkins University, Baltimore, MD 21218}\\
{\em $^3$Astrofi'sica da Universidade do Porto, Rua das Estrelas - 4150-762 Porto, Portugal}\\
{\em $^4$ Institut d'Astrophysique du CNRS, 98 bis Boulevard Arago, F-75014 Paris, France} \\
{\em $^5$ Steward Observatory, University of Arizona, 933 North Cherry Avenue, 
Tucson, AZ 85721, USA}\\
{\em $^6$ Apache Point Observatory, P.O. Box 59, Sunspot, NM 88349} \\

\vspace{1.5cm}

\begin {abstract}
We study  the recent                 
star formation histories of local galaxies by analyzing the scatter
in their colours and spectral properties. We present evidence that                  
the  distribution of star formation histories changes qualitatively above                  
a  characteristic stellar 
surface mass density of $3 \times 10^8 M_{\odot}$ kpc$^{-2}$, 
correponding to the transition between
disk-dominated (late-type) galaxies and bulge-dominated (early-type) systems.
When we average over subpopulations
of galaxies with  densities below this
value, we find that subpopulations 
 of all masses and densities form their stars at the same      
{\em average} rate per unit stellar mass.  However, 
the scatter in galaxy colours, stellar absorption line indices
and emission line strengths is larger for more compact galaxies of a given mass.  
This suggests that star formation occurs in shorter, 
higher amplitude events in galaxies with
smaller sizes.
Above the characteristic density, galaxy growth through star formation
shuts down  and the scatter in 
galaxy colours and spectral properties decreases.
We propose that in low density galaxies,
star formation events  are triggered               
when cold  gas is accreted onto a galaxy. 
We have used the new high resolution  Millennium
Simulation of structure formation in a ``concordance'' $\Lambda$CDM
Universe to quantify the incidence of these accretion events                
and we show that the observational data are well fit by a model
in which the consumption time of accreted  gas decreases with the surface
density of the galaxy  as $ t_{cons} \propto \mu_*^{-1}$.
The dark matter halos that host massive galaxies with high stellar 
surface mass densities are also
expected to grow through accretion, but the observations indicate
that in bulge-dominated galaxies, star formation is no longer coupled to 
the hierarchical build-up of these systems.                  
\end {abstract}
\end{center}
\vspace{1.5cm}
\normalsize
\pagebreak

\section {Introduction}

The physical  processes that regulate the
rate at which galaxies form stars and the timescale over
which galaxies evolve from star-forming to ``passive''
systems are still not understood. 
One key ingredient is the observed relationship between the large scale
star formation rate in a galaxy and the physical conditions
in its interstellar medium . Schmidt (1959) showed that 
the observed surface densities of gas and star formation in
external galaxies can be related by a power law,
\begin {equation} \Sigma_{\rm SFR} = A \Sigma^{N}_{\rm gas}. \end {equation}
The validity of the Schmidt law has been tested in many empirical
studies, with most measured values of $N$ falling in the range
1-2. The most comprehensive recent study is that of                
Kennicutt (1998), who used measurements of H$\alpha$, HI and CO
in both normal spiral galaxies and infrared-selected starburst 
galaxies to show that the disk-averaged star formation rates
in the combined sample could be represented by a Schmidt law with $N=1.4 \pm 0.15$
over several orders of magnitude in gas density.

If galaxies evolved in isolation as closed boxes, 
the Schmidt law implies that
their star formation rates  should decline smoothly with 
time. There is considerable evidence, however, that the
star formation rates in many galaxies have not been monotonic
with time, but have instead exhibited significant fluctuations. 
Analyses of the colour-magnitude diagrams of stars in Local Group
galaxies show that no two Local Group members have identical
star formation histories (see Grebel 2001; Dolphin et al 2005 for reviews).
Many of the dwarf galaxies appear to have formed their stars
in one or two discrete episodes or bursts. The Large and Small
Magellanic Clouds seem to have formed stars continuously, but
at a variable rate (Dolphin 2000; Smecker-Hane et al 2002; 
Harris \& Zaritksy 2004 ). An analysis of
solar neighbourhood data from the Hipparcos catalogue by Hernandez, Valls-Gabaud
\& Gilmore (2000) indicates that star formation in our own Milky
Way has also been variable (see also De la Fuente Marcos \& De la Fuente Marcos 2004). 
All these observations suggest that
star formation in galaxies is subject to either internal or
external {\em triggering mechanisms}.

It is not yet possible to derive accurate star formation histories
for galaxies outside our own Local Group. Instead, measurements of colours
and spectral features provide a ``snapshot'' of the ages of the stellar
populations and the current star formation rates in these systems.
An analysis of the intrinsic scatter in these measurements can diagnose
whether  star formation  has been continuous or bursty 
(Searle, Sargent \& Bagnuolo 1973).

Larson and Tinsley 
compared the scatter in the UBV colours of galaxies in the
Atlas of Peculiar Galaxies (Arp 1966) with ``normal'' galaxies
drawn from the Hubble Atlas of Galaxies (Sandage 1961). Their
results provided evidence for a ``burst'' mode of star formation
associated with violent dynamical phenomena.
Seiden \& Gerola (1979) and
Gerola, Seiden \& Schulman (1980) invoked     
internal interstellar medium processes to explain intermittent
star formation rates in dwarf galaxies. In their view, star formation
always occurs in discrete units with typical masses and sizes 
comparable to those of Giant Molecular Clouds. It was assumed that star  
formation in one unit could trigger star formation
in a neighbouring unit with a certain probability $P$. Once star formation
occurred, each unit required a certain  recovery period before
it was able to form stars again.  Because dwarf galaxies host only a            
small number of these units,  statistical fluctuations in the
number of units that are forming stars at any given time  result 
in integrated star formation histories that are bursty.
A key prediction of this scenario is that the lowest mass
systems experience the largest relative fluctuations in star formation rate.
Gerola et al did not compare the predictions of their models
with observations, arguing that the available samples were
incomplete, and in many cases heavily biased by selection effects.

In this paper, we revisit the question of the scatter among              
the recent star formation histories of local galaxies.
Two recent developments motivate this work. The first is
the availability of magnitude-limited samples of hundreds of thousands of
galaxies with high quality spectra and multi-band photometry. The second
development is the establishment of the cold dark matter (CDM) model, augmented with
a dark energy field, as the standard theoretical paradigm for structure
and galaxy formation. Measurements of the cosmic microwave background
by the WMAP satellite (Spergel et al 2003) and the galaxy power spectrum by the 2dFGRS and 
Sloan Digital Sky Surveys (Percival et al 2001; Tegmark et al 2004) 
 have boosted confidence in the validity
of the model and have allowed precise determination of the geometry
and matter content of the Universe. The collapse of the fluctuations
in the dark matter component and the subsequent build-up of structure
can be accurately computed using direct numerical simulation.
This allows the incidence of mergers and of accretion onto
galaxies to be predicted  in detail as a function of cosmic epoch.

Our paper is divided into two parts. In the first part, we analyze
a large sample of galaxies drawn from the Sloan Digital Sky Survey. 
We study how the scatter in galaxy colours and age-dependent stellar absorption
features such as the 4000 \AA\ break
depend on  the mass and on the
structural properties of galaxies. We demonstrate that the            
galaxy-to-galaxy scatter  does not depend primarily on the
mass of the galaxy, but rather on structural properties such as surface density
and concentration. The scatter increases monotonically
as function of these parameters  and reaches 
a {\em maximum}, before dropping again. We identify the location of
this maximum as the point where the dominant population component transitions from     
disk-dominated (late-type) galaxies to bulge-dominated (early-type) systems.
We also study how the distributions of star formation rates  deduced from
emission lines in the spectra  
change between diffuse, low surface density galaxies
and compact, more concentrated systems of the same stellar mass.

In the second part of the paper, we introduce a simple model that can 
explain many of the  observational trends in a self-consistent way.
In our scenario, star formation in late-type, disk-dominated  galaxies is 
regulated by two factors: a) the infall of gas that occurs as dark matter halos 
assemble through the process of hierarchical clustering, b) the time for the conversion
of this gas into stars, which is determined by the density of the system.
The infall rates are calculated using a new high resolution
simulation of structure formation in a $\Lambda$CDM ``concordance'' cosmology.
We show that the inferred gas conversion times scale with the surface
density of the galaxy approximately as                            
predicted by the Schmidt/Kennicutt
law of star formation. 

\section {The Sample}

The data analyzed in this study are drawn from the Sloan Digital Sky
Survey (SDSS).  The survey goals are to obtain photometry of a quarter
of the sky and spectra of nearly one million objects.  Imaging is
obtained in the {\em u, g, r, i, z} bands (Fukugita et al 1996;
Smith et al 2002; Ivezic et al 2004) with a special purpose drift scan camera
(Gunn et al 1998) mounted on the SDSS 2.5~meter telescope (Gunn et al 2005) at
Apache Point Observatory.  The imaging data are photometrically
(Hogg et al 2001; Tucker et al 2005) and astrometrically (Pier et al 2003)
calibrated, and used to select stars, galaxies, and quasars for
follow-up fibre spectroscopy.  Spectroscopic fibres are assigned to
objects on the sky using an efficient tiling algorithm designed to
optimize completeness (Blanton et al 2003a).  The details of the
survey strategy can be found in (York et al 2000) and an
overview of the data pipelines and products is provided in the Early
Data Release paper (Stoughton et al 2002).

Our parent sample for this study is composed of 397,344 objects which
have been spectroscopically confirmed as galaxies and have data
publicly available in the SDSS Data Release~4
(Adelman-McCarthy et al  2005).  These galaxies are part of the
SDSS `main' galaxy sample used for large scale structure studies
(Strauss et al 2002) and have Petrosian $r$ magnitudes in the
range $14.5 < r < 17.77$ after correction for foreground galactic
extinction using the reddening maps of
Schlegel,Finkbeiner \& Davis (1998).  Their redshift
distribution extends from $\sim0.005$ to 0.30, with a median $z$ of 0.10.

The SDSS spectra are obtained with two 320-fibre spectrographs mounted on the SDSS 
2.5-meter telescope.  Fibers 3 arcsec in diameter are manually plugged
into custom-drilled aluminum plates mounted at the focal plane of the
telescope. The spectra are exposed for 45 minutes or until a fiducial
signal-to-noise (S/N) is reached.  The median S/N per pixel for
galaxies in the main sample is $\sim14$.  The spectra are processed by
an automated pipeline,  which flux and
wavelength calibrates the data from 3800 to 9200~\AA.  The
instrumental resolution is R~$\equiv \lambda/\delta\lambda$ = 1850 --
2200 (FWHM$\sim2.4$~\AA\ at 5000~\AA). At the median 
redshift of the SDSS main galaxy sample ($z\sim0.1$) the target
galaxies subtend several arcseconds on the sky.  The fraction of
galaxy light falling in the 3 arcsec fibre aperture is typically
about 1/3. 

In this paper we use the amplitude of the 4000 \AA\ break 
(the narrow version of the index defined in
Balogh et al. 1999)
and the strength of the H$\delta$ absorption line (the Lick
$H\delta_A$ index of Worthey \& Ottaviani 1997) as diagnostics of the
stellar populations of the host galaxies. Both indices
are corrected for the observed contributions of the emission-lines
in their bandpasses (see Tremonti et al 2004 for a more detailed
discussion).
As described in Kauffmann et al (2003a), a  library of
32,000 model star formation histories 
and the measured $D_n(4000)$ and $H\delta_A$ indices are used
to obtain a median likelihood  estimate of  the 
$z$-band mass-to-light ratio for each galaxy.
By comparing the colour predicted by the  best-fit model to 
the observed colour of the galaxy,
we also estimate the attenuation of the starlight due to dust.

The SDSS imaging data provide the basic structural parameters 
that are used in this analysis.
We use the $z$-band as our fiducial filter because it is the least sensitive to the
effects of dust attenuation.
The $z$-band absolute magnitude, combined with our 
estimated values of M/L and dust attenuation
$A_z$ yield the stellar mass ($M_*$). The half-light radius in the $z$-band and the
stellar mass yield the effective stellar surface mass-density
($\mu_* = M_*/2\pi r_{50,z}^2$). As a proxy for Hubble type we use
the SDSS ``concentration'' parameter $C$, which is defined as the ratio
of the radii enclosing 90\% and 50\% of the galaxy light in the $r$ band
(see Stoughton et al. 2002). Strateva et al. (2001) find that galaxies
with $C >$ 2.6 are mostly early-type galaxies, whereas spirals and irregulars
have 2.0 $< C <$ 2.6.

The procedures we adopt for estimating stellar masses
are described in detail in Kauffmann et al (2003a). The relations between
stellar absorption line indices, stellar masses and structural
parameters are discussed in Kauffmann et al (2003b).
We have also used star formation rates estimated from the emission
lines using the methods described in Brinchmann et al (2004). 
The parameters from our analyses have been made publically available 
at http://www.mpa-garching.mpg.de/SDSS/.
Where appropriate
we adopt a cosmology of $\Omega = 0.3$, 
$\Omega_{\Lambda} = 0.7$, and H$_0$ = 70 km~s$^{-1}$ Mpc$^{-1}$.

\section {Observational Analysis}

\subsection {Analysis of Matched Galaxy Pairs}

In order to draw meaningful physical conclusions
from an analysis of the scatter in  the colours or spectral features of
galaxies, it is important that the galaxy sample be as homogeneous as possible.
The mean stellar ages of
galaxies are known to correlate strongly with properties such as galaxy
luminosity, mass, concentration and size (Tinsley 1968). The reader is referred to
Kauffmann et al (2003b) and Blanton et al (2003b) for
a detailed analysis of these correlations for galaxies in the SDSS.
Throughout this paper we analyze 
galaxies that are selected to be as similar
as possible in terms of their stellar masses and structural properties.
In order to minimize any additional scatter in spectroscopic
parameters induced by mismatched apertures, we also require
that the galaxies be closely matched in redshift. 

In this section, we analyze a sample of matched galaxy {\em pairs}.
In our default sample, pairs are selected to have $\Delta \log M_* < 0.05$, 
$\Delta \log \mu_* < 0.05$, $\Delta C < 0.05$, 
$\Delta \sigma < 15$ km s$^{-1}$ and $\Delta z < 0.005$ (where
$M*$ and $\mu_*$ are in units of $M_{\odot}$ and $M_{\odot}$ kpc$^{-2}$,
$C$ is the concentration index, $\sigma$ is the stellar velocity dispersion
and $z$ is the redshift).
For galaxies with stellar masses less than $10^{10} M_{\odot}$
the tolerances are relaxed to 
$\Delta \log \mu_* < 0.1$, $\Delta C < 0.1$, 
$\Delta \sigma < 25$ km s$^{-1}$ and $\Delta z < 0.01$ because there
are fewer low mass galaxies in our sample. Even with these very
stringent matching constraints, we find 139,676 distinct
galaxy pairs (where `distinct' means that no galaxy
is assigned to a pair more than once). We now use this pair sample
to study how the variation in star formation history between structurally
similar galaxies depends on properties such as mass, concentration
and stellar surface mass density. 

Since we wish to analyze the properties of our sample
as a function of  stellar mass rather than of
$r$-band absolute magnitude, it is important to ensure that
survey selection effects do not bias our estimates of the scatter,
because at fixed stellar mass,  young galaxies with small mass-to-light ratios 
are detected out to greater distances
than old galaxies with large mass-to-light ratios. For each stellar mass,
we calculate the limiting  redshift out to which a galaxy with
maximal $M/L$ is detected in the survey. The maximal
$M/L$ is computed using the Bruzual \& Charlot (2003) population
synthesis models assuming a single star formation burst at $z=10$.   
Any pair with a redshift greater than this limit is excluded
from the analysis.

Before we analyze our default sample, it is useful to illustrate
the effect of increasing the precision with which galaxies are paired
by increasing the number of parameters
which are required to match within small tolerances.
This is illustrated in Figure 1 where each panel corresponds to pairs
matched in a different number of parameters. For each panel we rank        
all the pairs in order of increasing stellar 
mass and then divide  them into
groups that each contain 1500 pairs. 
For each group, we calculate  the r.m.s.  difference
in the 4000 \AA\ break strengths of the paired galaxies.   
\footnote {There are two reasons why D$_n$(4000) is used as our default stellar age
indicator throughout this paper: 1) It has very small measurement errors;
2)It is more insensitive to dust than galaxy colours.}   
We also correct for the contribution of observational errors  
by subtracting in quadrature the errors in $\Delta$D$_n$(4000) for each pair.
From now on, we will refer to this quantity as the {\em variation} in D$_n$(4000).
Because there are always a fixed number of pairs in each group, the
error on each point plotted in Figure 1 should be similar.
It can be estimated  from the scatter in the plotted 
points at intermediate values
of $M_*$, where the number of pairs is very large.

The top left panel shows the dependence
of the variation  on $M_*$ if pairs are only matched in stellar
mass and redshift. As can be seen, the variation peaks
at $\log M_* \sim 10.2$. As discussed in Kauffmann et al (2003b),
most low mass galaxies have young stellar populations
and most high mass galaxies have old stellar populations.
A sharp transition between these two regimes
occurs at a characteristic stellar mass of $3 \times 10^{10} M_{\odot}$
(also at a characteristic stellar surface mass density 
of $3 \times 10^{8} M_{\odot}$ kpc$^{-2}$ and a concentration index of 2.6).
This explains why the variation peaks 
around this stellar mass.  However, the top right and bottom left panels 
show that this
peak is largely suppressed  once galaxies are matched  in both stellar
mass and  structural parameters such as size or concentration.
The bottom right panel shows the variation  as a function of stellar mass
for our default pair sample, which is matched in redshift,  $M_*$, size, concentration
and velocity dispersion.

\begin{figure}
\centerline{
\epsfxsize=14cm \epsfbox{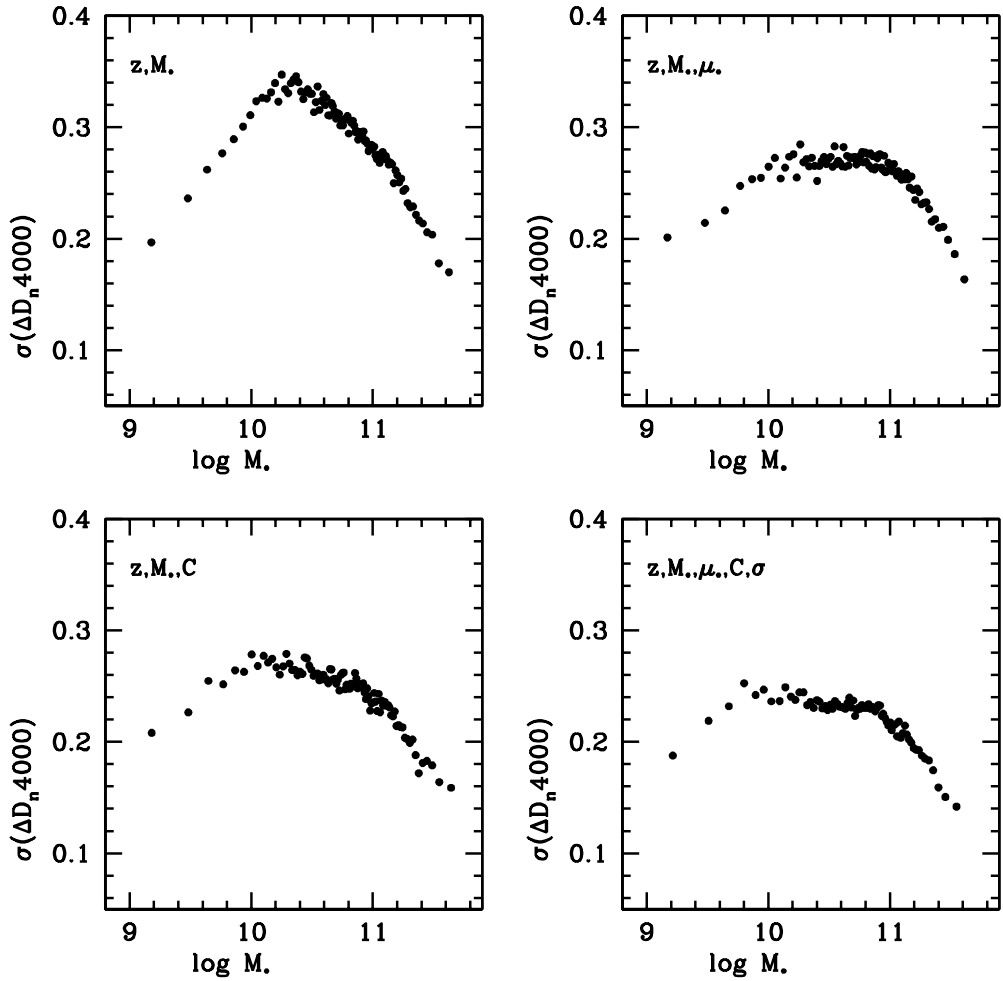}
}
\caption{\label{fig1}
\small
The r.m.s. difference in D$_n$(4000) is plotted as a function of galaxy
mass for the pair sample (see text for details). The different panels
show the effect of matching pairs in successively larger numbers of
parameters. Note that D$_n$(4000) ranges from 1.2 to 2.4 for SDSS galaxies.}
\end {figure}
\normalsize

In Figure 2, we plot the variation  in D$_n$(4000) as a function
$M_*$ , $\mu_*$,  $C$ and $\sigma$ for our default pair sample. Interestingly
the peak in the variation is larger and also more
sharply defined  for the parameters $\mu_*$ and $C$ 
than for $M_*$ and $\sigma$.
For the two structural
parameters, the peak  occurs at  $C=2.6$ $\log \mu_*=8.5$, very close 
to the characteristic values 
given in  Kauffmann et al (2003b). The smallest
variations in D$_n$(4000), on the other hand,  are obtained for galaxies with the
largest velocity dispersions.
The peak in the variation is interesting because it indicates
that there are certain values of the structural parameters
where we are least able to predict the age
of the stellar populations in a galaxy. This shows  that the {\em transition}
from star-forming  to ``passive'' is most likely driven by
the structure of the galaxy rather than its mass.    

We now demonstrate that the peak in variation  
at $C$ around 2.6 and $\log \mu_* = 3 \times 10^8 M_{\odot}$ 
is not confined to one  particular
stellar indicator. Figure 3  compares the
variations as a function of $M_*$ and $C$ for three different
age-sensitive indicators: the 4000 \AA\ break (as before), 
the $g-r$ Petrosian and fibre colours,
and the H$\delta_A$ Lick index. As can be seen, the
maximum at $C=2.6$ is apparent for all three indicators. (We note that similar
results are found as a function of $\mu_*$, with the peak
always at $\mu_* \sim 3 \times 10^8 M_{\odot}$ kpc$^{-2}$.) The variation in D$_n$(4000)
and H$\delta_A$ exhibit very similar behaviour as a function of
$M_*$ and $C$. The variation in colour exhibits somewhat different behaviour,  
particularly for the lowest mass galaxies where it rises rather than falls.
The variation in colour also falls off less steeply 
at the highest stellar masses.
The results presented in the  middle panel of Figure 3 show that
we obtain almost identical results for both fibre colours (which, like the indices,
sample the light from  the central 3 arcsec of the galaxy)  and Petrosian
colours (which sample the light from the entire galaxy). This suggests
that the differences between colours and indices are not caused by the fact that the
two quantities are measured within different apertures. 
Galaxy colours are sensitive to dust extinction as well as to
mean stellar age and it is thus likely that variable extinction in different galaxies
contributes significantly to the variation in colour.

We now consider how much of the variation is contributed by the dependence of
the colours and star formation rates of galaxies on their environment. 
As described by  numerous authors (see Hogg et al 2003; Kauffmann et al 2004;
Blanton et al 2005 for discussions pertaining to SDSS galaxies), galaxy colours
are strongly dependent on local density, with the fraction of red
galaxies increasing strongly in the highest density environments.
Following Kauffmann et al 2004, we have calculated galaxy counts in
cylindrical apertures  of radius 2 Mpc and depth $\pm$ 500 km s$^{-1}$ around each
galaxy. We repeat our analysis by selecting pairs from a subsample that 
contains  half the galaxies 
in the lowest density environments.                                      
The results are shown as blue symbols in Figure 4. Also shown are 
results for a subsample that 
includes 20\% of the galaxies in the densest regions (red symbols).
As can be seen, if galaxies are restricted to lie in more similar
environments, the overall amplitude of the variation decreases, although
the reduction is small. \footnote{ As shown in Figure 7 of Kauffmann
et al 2004, the largest shifts in the median value of D$_n$(4000) as
a function of $C$ and $\mu_*$  occur for the two densest (red and magenta)
bins, which together comprise less than 10\% of all the galaxies in the sample.
Because the distribution of
D$_n$(4000) is {\em bimodal}, 
the variation in  D$_n$(4000) is also expected to be less sensitive
to environment than the median value.}
However,
the trends as a function of $C$ and $\mu_*$ remain the same. 

Finally, in Figures 5 and 6, we plot the variation in D$_n$(4000) for
six narrow ranges in $\log M_*$. Low mass galaxies have lower concentrations
and stellar surface mass densities than high mass galaxies. Nevertheless,
our results show that the trend for the variation to increase
with concentration (surface density) for $C<2.5$ ($\log \mu_* < 8.5$)
and to decrease with concentration (surface density) for $C>2.5$
($\log \mu_*> 8.5$) occurs for all galaxies {\em independent of
their stellar mass}. It is curious that at low stellar surface
densities, lower mass galaxies have larger variation  than higher   
mass galaxies, but at low concentrations, the opposite appears to be true.
At high surface densities and concentrations, the most massive galaxies
always have the smallest variation.
  
\begin{figure}
\centerline{
\epsfxsize=14cm \epsfbox{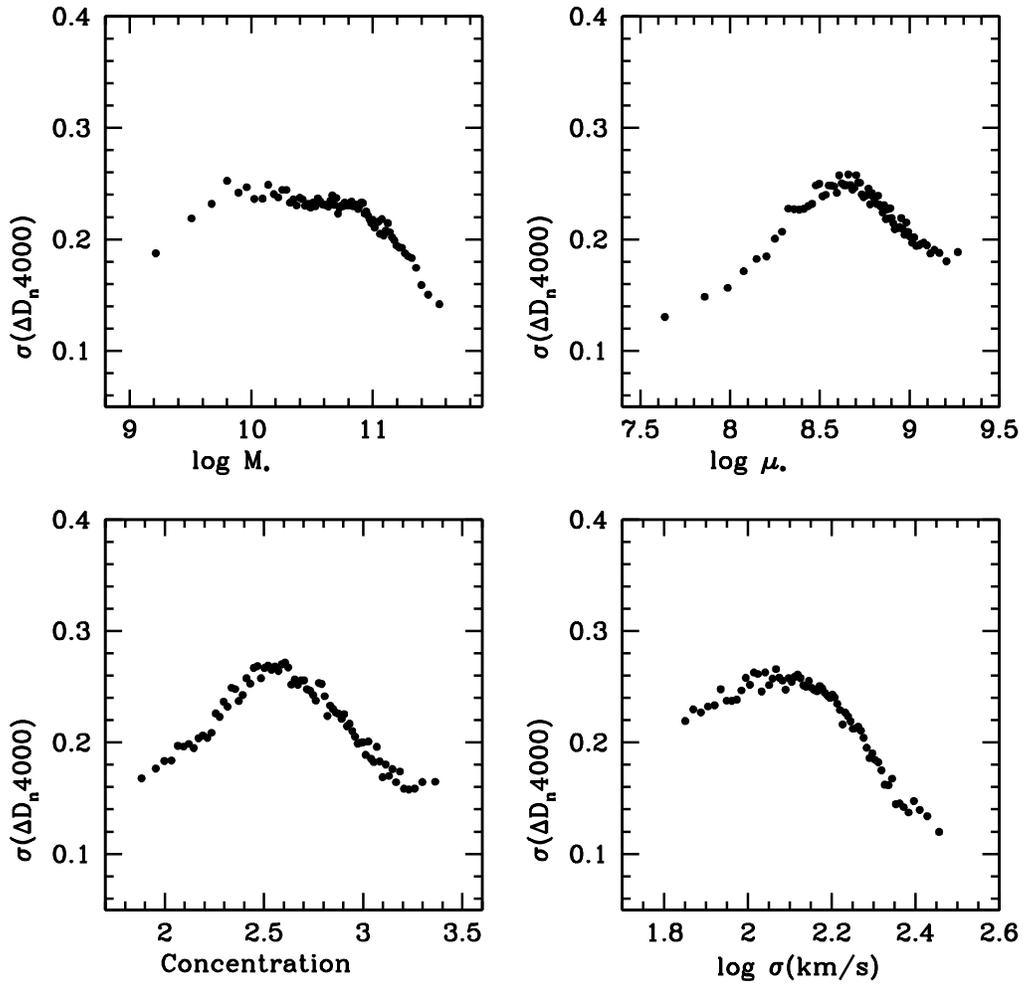}
}
\caption{\label{fig2}
\small
The r.m.s. difference in D$_n$(4000) is plotted as a function of 
stellar mass, stellar surface mass density, concentration
and velocity dispersion for
a galaxy pair sample that has been matched according to
redshift, stellar mass, stellar surface density, concentration,
and velocity dispersion.} 
\end {figure}
\normalsize

\begin{figure}
\centerline{
\epsfxsize=16cm \epsfbox{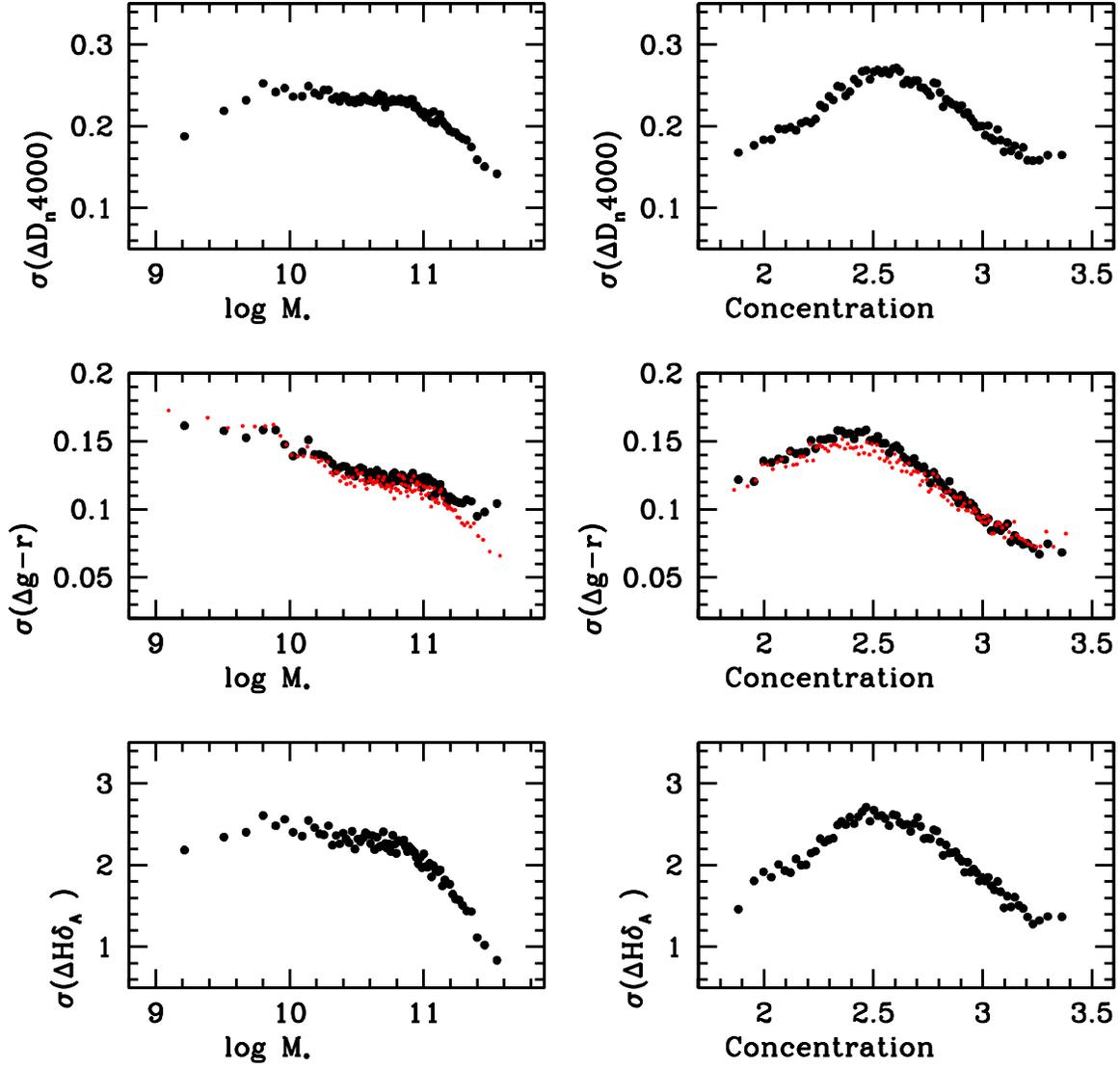}
}
\caption{\label{fig3}
\small
The r.m.s. difference in D$_n$(4000), $g-r$ Petrosian (black points)
and fibre (red points) colours,
and  H$\delta_A$ is plotted as a function of 
stellar mass and concentration
for a galaxy pair sample matched in $z$,$M_*$,$\mu_*$, $C$, and $\sigma$. 
Note that $g-r$ colors span the range 0.2-1.1 (at z=0.1). H$\delta_A$
spans the range -3.5---7 \AA\ and D$_n$(4000) the range 1.1-2.4.}
\end {figure}
\normalsize

\begin{figure}
\centerline{
\epsfxsize=8.5cm \epsfbox{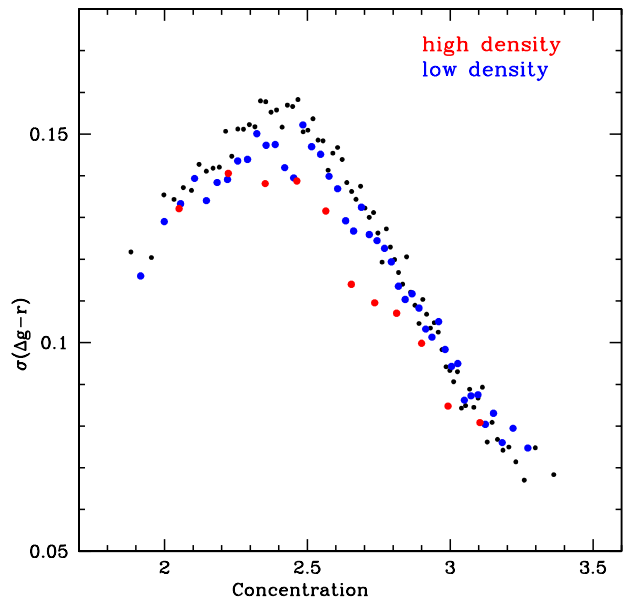}
}
\caption{\label{fig4}
\small
The r.m.s. difference in $g-r$ Petrosian 
colour is ploted as a function of concentration for 
the whole sample (black points), for 50\% of the galaxies
in the lowest density environments (blue points) and for
20\% of galaxies in the highest density environments (red points).
The galaxy pair sample has been  matched in $z$,$M_*$,$\mu_*$, $C$, and $\sigma$.} 
\end {figure}
\normalsize

\begin{figure}
\centerline{
\epsfxsize=8.5cm \epsfbox{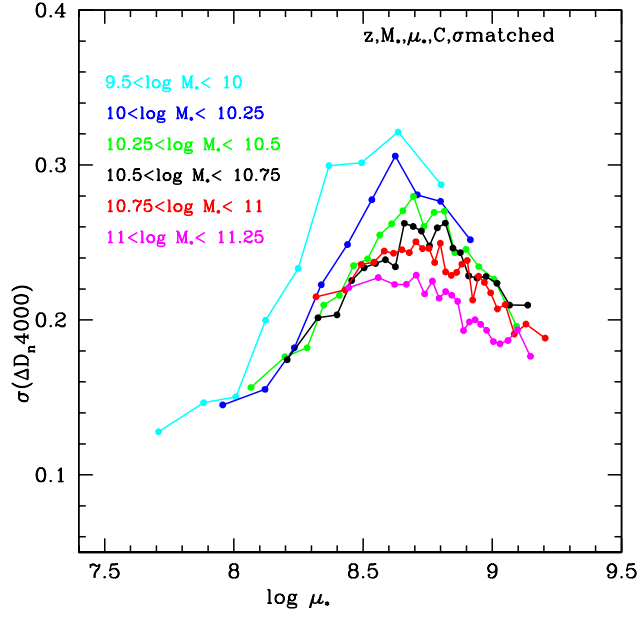}
}
\caption{\label{fig5}
\small
The r.m.s. difference in D$_n$(4000) is plotted as a function of 
stellar surface mass density for 6 ranges in stellar mass. 
The galaxy pair sample has been  matched in $z$,$M_*$,$\mu_*$, $C$, and $\sigma$.} 
\end {figure}
\normalsize

\begin{figure}
\centerline{
\epsfxsize=8.5cm \epsfbox{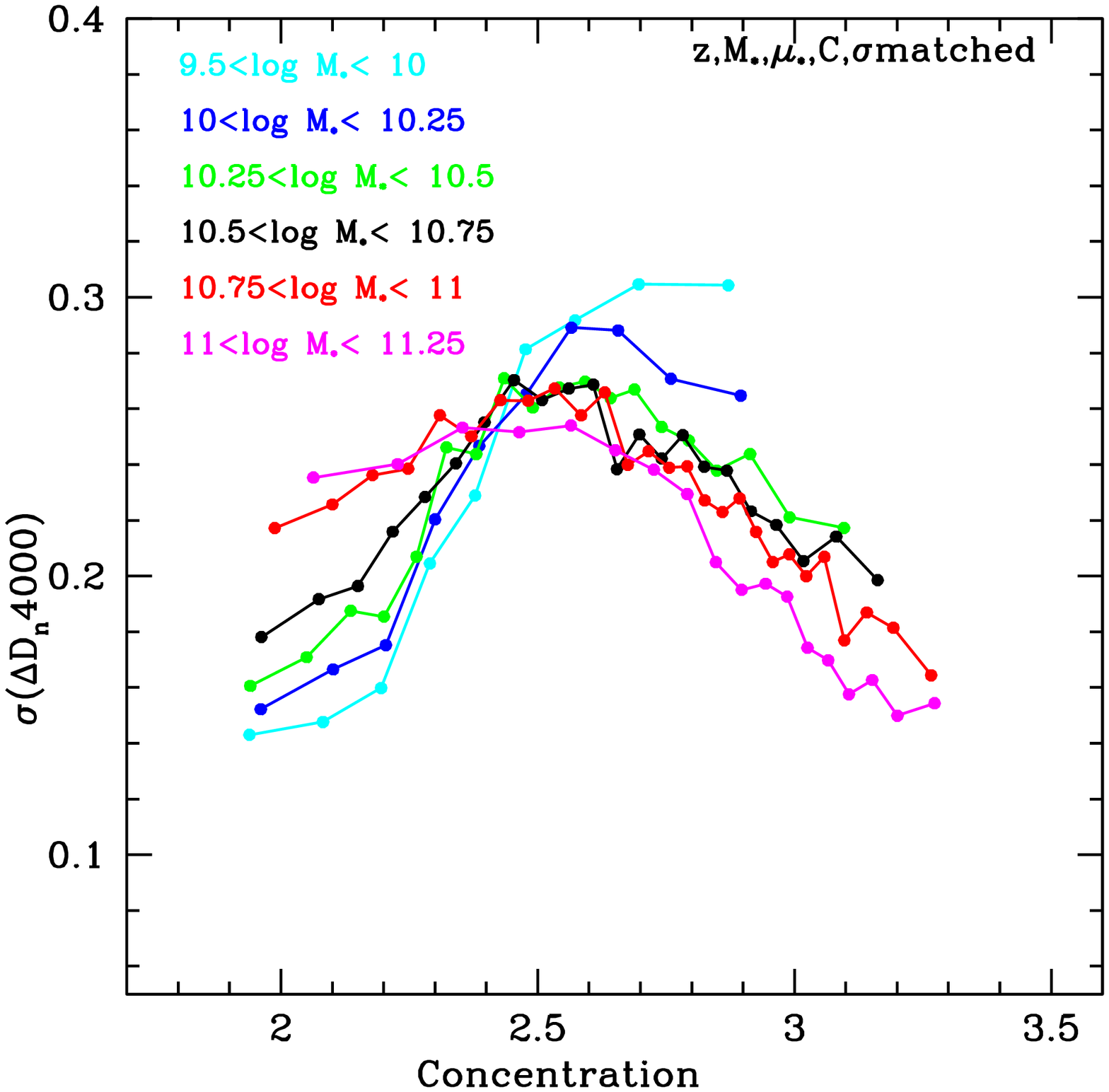}
}
\caption{\label{fig6}
\small
The r.m.s. difference in D$_n$(4000) is plotted as a function of 
concentration for 6 ranges in stellar mass. 
The galaxy pair sample has been  matched in $z$,$M_*$,$\mu_*$, $C$, and $\sigma$.} 
\end {figure}
\normalsize

\subsection {Distribution Functions}

A more intuitive way of phrasing  the results presented in the
previous section is in terms of the ability to 
{\em predict} the stellar populations
of a galaxy, given full information about its mass and structural
properties. The results presented in Figures 3-6 show that
the stellar populations are least predictable at C=2.5
and $\mu_* = 3 \times 10^8 M_{\odot}$ and most predictable 
both at very low and at very high values of $\mu_*$ and $C$.

High values of $\mu_*$ and $C$ correspond to the regime of 
galaxy spheroids and  bulges.  ``Predictability'' simply
reflects the fact that star formation has switched
off in bulge-dominated  galaxies  and their stars are now uniformly old.
More puzzling is the increase in variation  as a function
of $\mu_*$ and $C$ for the star-forming, disk-dominated galaxies in our sample.
In this section, we attempt to gain more insight into this trend.

As shown in Figure 1,  matching galaxies in redshift, stellar mass and a single   
structural parameter (either concentration or surface density)
already causes the variation in D$_n$(4000) to decrease to close to its
minimum level. In this section, we 
present  the  distribution in D$_n$(4000) of
galaxies in narrow intervals of mass and stellar surface mass density.    
This is illustrated  in Figures 7 and 8  for galaxies with stellar masses in the ranges
$2-5 \times 10^9 M_{\odot}$ and $2-5 \times 10^{10} M_{\odot}$.
In order to make these plots, we have have relaxed the redshift matching 
tolerance to $\Delta z < 0.01$ and instead of
matching pairs,  we find {\em all}  galaxies that
lie within the specified range in $M_*$ and $\mu_*$.
Distributions are shown for  samples  containing at least 700 galaxies.

At stellar surface mass densities less than  
$3 \times 10^8 M_{\odot}$ kpc$^{-2}$, the main body
of the population has  D$_n$(4000) values peaked at around 1.3-1.4. 
There is a tail of objects with higher values of D$_n$(4000)  and the fraction of galaxies
contained in this tail increases with stellar surface mass  density.
At stellar surface mass densities greater than  
$3 \times 10^8 M_{\odot}$ kpc$^{-2}$, the parity of the 
distribution rapidly switches and
the main body of the population is peaked at large values
of D$_n$(4000), with a tail of objects towards the blue.

\begin{figure}
\centerline{
\epsfxsize=15cm \epsfbox{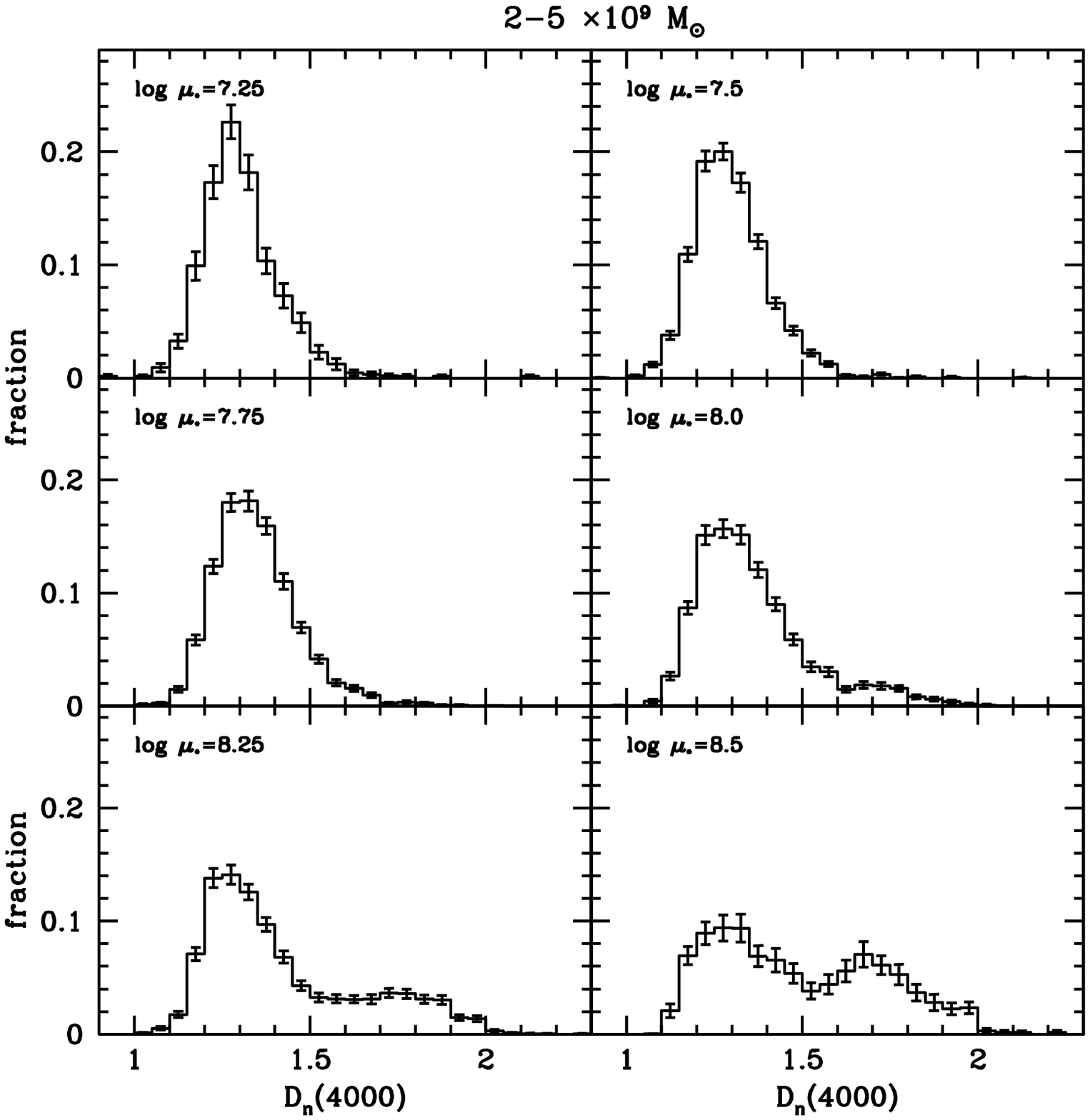}
}
\caption{\label{fig7}
\small
The distribution of D$_n$(4000) is plotted  
for 6 ranges in stellar surface mass density for galaxies with
stellar masses in the range $2-5 \times 10^9 M_{\odot}$. Error bars
have been computed using a standard bootstrap resampling technique.} 
\end {figure}
\normalsize

\begin{figure}
\centerline{
\epsfxsize=15cm \epsfbox{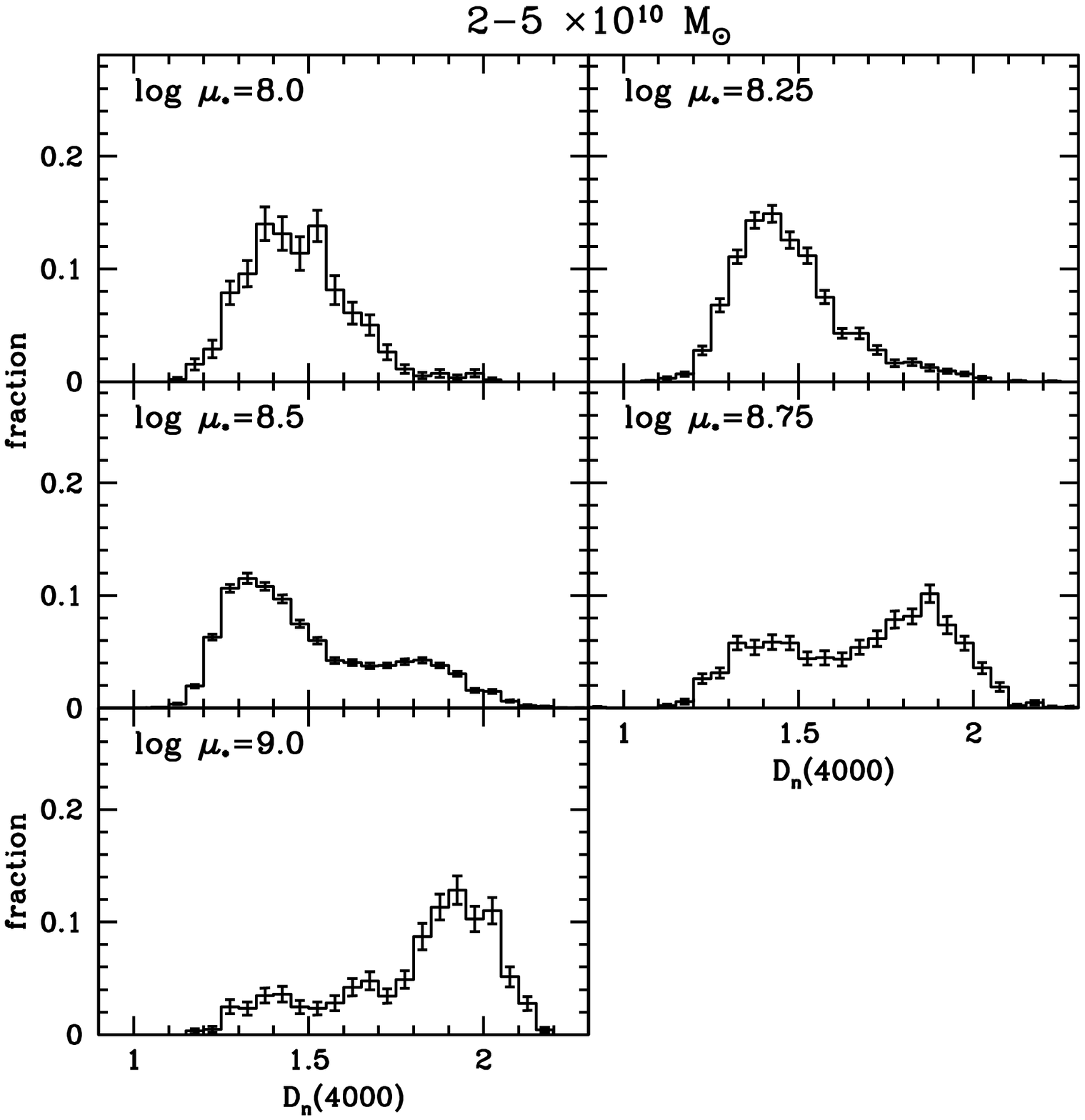}
}
\caption{\label{fig8}
\small
As in Figure 7, except for galaxies with   
stellar masses in the range $2-5 \times 10^{10} M_{\odot}$. Error bars
have been computed using a standard bootstrap resampling technique.} 
\end {figure}
\normalsize

There are two  scenarios that come to mind  when attempting to  explain
these trends:
\begin {enumerate}
\item As surface density increases, an increasing fraction of galaxies
have stopped forming stars. In this scenario, the blue and red populations are
disjoint. Once a galaxy stops forming stars, it joins the red population
and stays there permanently.                    
\item As surface density increases, galaxies still form the same
total number of stars, but  the characteristic {\em timescale} over
which star-forming events occur becomes shorter. In this scenario, the blue
and red populations are not disjoint. Galaxies may stop forming stars temporarily,
but star formation will be triggered again at a later time. Galaxies 
thus migrate from the blue peak to the red and then back again.  The red
population is more dominant at high surface densities, because the
characteristic dynamical times in these objects are smaller and
star formation occurs in shorter bursts of higher intensity.
Following the end of a burst of star formation, D$_n$(4000)    
increases to values characteristic of old stellar populations in timescales
of less than a Gyr for bursts involving a small fraction of the total mass
of the galaxy   
(Charlot \& Silk 1994).  

\end {enumerate}

These two scenarios can be distinguished by a very simple test. In scenario 1,
the {\em average} amount of star formation for the population as a whole
should decrease as the surface density becomes larger and as more
and more  galaxies ``switch off''. In scenario 2, the average amount of star formation
should remain constant as a function of surface density. Because star
formation occurs in shorter, more intense bursts, one also expects to see a larger
fraction of  galaxies with strong emission lines.

The results of this test are shown in Figure 9. We have used the in-fibre  
specific star formation
rates estimated  by Brinchmann et al (2004) to calculate the  average
star formation rate per unit stellar mass  as a function of stellar surface
mass density. The results  are shown as thick solid lines
on Figure 9 for nine  different ranges in $\log M_*$. The dotted curves indicate
the upper 95th percentile and lower 25th percentile of the {\em distributions}
of $\log$ SFR$/M_*$ at a given value of $\log \mu_*$. 

As can be seen, the  
average specific star formation rate  $\overline{\rm SFR}/M_*$ remains remarkably constant for
low mass galaxies. For higher mass galaxies,
$\overline{\rm SFR}/M_*$ remains 
 constant for $\log \mu_* < 8.5$ and decreases
at higher surface densities. \footnote {Note that the timescale associated
with this specific star formation rate $(\overline{SFR}/M_*)^{-1}
\simeq 6 \times 10^9$ years. For our assumed Kroupa (2001) IMF, the return
fraction is $\sim 0.5$, so this corresponds to
a mean timescale for the build-up of the galaxies of
$10^{10}$ years. The satisfying consistency with the Hubble time shows
the consistency of the calibrations of our observational
stellar mass and star formation rate indicators. We have also checked that
we obtain the same mean time scale when we use the aperture-corrected
specific star formation rates derived by Brinchmann et al (2004).}  Over the
range of surface densities where $\overline{\rm SFR}/M_*$ is
constant , the upper 95th percentile of the distribution of
$\log$ SFR/$M_*$ exhibits a small but significant increase,
while the lower 25th percentile drops very steeply.
This shows that even though the  specific star formation
rates averaged over the  galaxy population as a whole do not depend  
on density or on mass for $\log \mu_* < 8.5$, the star formation histories
of the galaxies in our sample do change systematically as
$\mu_*$ increases. At a given stellar mass, more compact and dense galaxies
experience stronger  bursts as well as extended periods of inactivity.
We conclude that Scenario 2
provides a good  description of what is happening below the characteristic surface
density. Above this density, the decrease both in the average value of SFR/$M_*$
and in the upper and lower percentiles indicates that star formation is
truly shutting down at very high densities and that Scenario 1 provides a more accurate
description of the current state of these galaxies.
Figure 9 also shows that the total spread in SFR/$M_*$ reaches a {\em maximum} at the
characteristic surface density . This is in good agreement with the results shown
in the previous section. 

Finally we note that the average value of SFR/$M_*$
below the characteristic surface density is almost      
{\em independent of stellar mass}.
Over the mass range from $10^9 M_{\odot}$ to $5 \times 10^{10} M_{\odot}$,
$\log \overline{\rm SFR}/M_*$ always has values in the range -9.7 -- -9.9.
This ``universality'' in the value of the average specific star
formation rate was also discussed in Brinchmann et al (2004).
It is quite remarkable and we will
try to provide an explanation for why this should be the case 
in the following section.

\begin{figure}
\centerline{
\epsfxsize=16cm \epsfbox{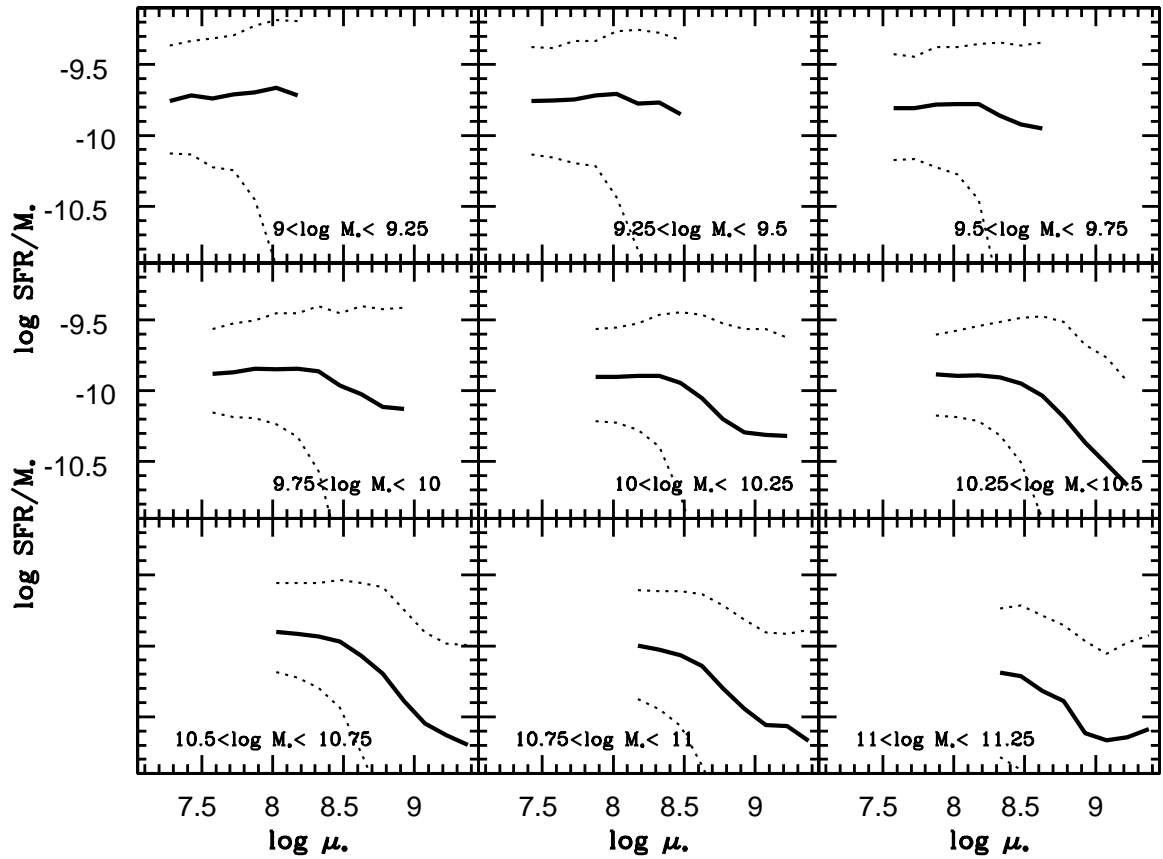}
}
\caption{\label{fig9}
\small
The thick solid curves indicate the average star formation rate divided by
the stellar mass within the fibre aperture. The lower and upper dotted curves
show the 25th and 95th percentiles of the distribution of $\log$ SFR/$M_*$.
Results are shown for galaxies in nine ranges of $\log M_*$.}
\end {figure}
\normalsize

\section {A Simple Model}

\subsection {The Millennium Run Simulation}
In the previous section we showed that at 
stellar surface densitities below $\sim 3 \times 10^8 M_{\odot}$ kpc$^{-2}$,
the mean specific star formation rate of galaxies of given mass and surface density does
not depend on either mass or surface density, but the scatter
in star formation rate between galaxies of the same mass and
surface density increases with surface density. The simplest interpretation
is then that all these galaxies form stars at the same time-averaged rate, but
that the star formation is burstier in high surface density galaxies. 
The question that then arises is what is reponsible for triggering
episodic star formation in these galaxies?

In hierarchical models of galaxy formation (e.g. White \& Rees 1978; White \& Frenk 1991;
Kauffmann, White \& Guiderdoni 1993; Cole et al 1994; Kauffmann
et al 1999; Somerville \& Primack 1999; Cole et al 2000; Croton et al 2005), galaxies
grow by merging and by  accretion of gas from their surroundings.
At early times and in low mass structures, the
gas cooling times are short. The supply of gas onto a galaxy is
regulated by infall of new material that occurs as its surrounding dark matter
halo grows with time. At late times and in massive structures, the cooling
times become longer than the local  dynamical time. 
Infalling gas  then shock heats to the virial
temperature of the halo and forms a hot atmosphere which cools quasi-statically. 
In recent work, Forcada-Miro \& White (1997) and  Birnboim \&     
Dekel (2003)  showed that for realistic dark matter halos
in a $\Lambda$CDM cosmology,  the transition between the 
infall-regulated and cooling flow regimes occurs at a halo
mass of approximately $10^{12} M_{\odot}$ at the present day. 
Thus, the majority of lower
mass galaxies in our sample  
may be experiencing infall-regulated gas accretion.

In  early work  (e.g. White \& Frenk 1991)
the infall-regulated regime was modelled in an approximate way -- every
halo was assumed to have the same  accretion history. With the advent
of large  N-body simulations, it has become possible to follow the 
build-up of dark matter halos of  Milky Way mass and larger in cosmologically
interesting volumes   
(Kauffmann et al 1999).
In this paper we make use of a new high-resolution simulation --
the {\em Millennium Simulation} -- recently carried out by the Virgo
Consortium and described in  Springel et al (2005).
With 2160$^3$ particles in a cubic region 500$h^{-1}$ Mpc on a side
(i.e. a particle mass of $8.6 \times 10^8 h^{-1} M_{\odot}$),
the simulation offers an unprecedented combination of high spatial
resolution and a large simulated volume and it permits detailed statistical 
analysis of the assembly histories of
dark matter halos down to masses typical of LMC/SMC-type galaxies. 

The post-processing of the simulation data includes the identification
of gravitationally bound dark matter subhalos orbiting within larger
virialized structures. This is carried out using an extended version of the
SUBFIND algorithm described in Springel et al (2001). These subhalos
are then tracked over time in order to construct hierarchical
merging trees that describe in detail how cosmic structures build up
over time. These merging trees are stored at 64 different redshifts
spaced logarithmically from $z=127$ to $z=0$ and they form
the basis for the analysis carried
out in this section.

\subsection {Mass accretion histories for dark matter halos}

The left panel of Figure 10 shows the {\em average} mass accretion 
histories for dark matter halos 
in a number of different mass ranges. A galaxy with the mass
of the Milky Way would be expected to reside in a dark matter halo
of mass of a few $\times 10^{12} M_{\odot}$ at the present day,
so our accretion histories are relevant for the
mass range of galaxies contained in our
observational samples. To calculate the average history, we track
the mass of the most massive progenitor of each halo back in time, i.e.
at time $t_i$, we find the largest progenitor of the object that is the               
largest progenitor at time $t_{i+1}$ (where $t_{i+1}> t_i$, and $t$
measures the time since the Big Bang).
This procedure is repeated for all the halos in the simulation
volume and the average histories are calculated by summing over all the
halos in the given mass range.
Note that this procedure does not ensure that the object identified
at time $t_{i}$ is the most massive among {\em all possible} progenitors
of the halo at that time (i.e. when the trunk of the merger tree
bifurcates, the largest branch at the point of bifurcation may not lead to 
the largest branches at all earlier epochs).  In this analysis we are   
mainly concerned with the {\em recent} accretion histories of  galaxies,
typically before any bifurcation has taken place, so these subtleties 
will not be important for us.

Figure 10 indicates that more massive halos assemble later than less
massive halos.  Over the range of masses shown in the plot, this effect is
rather weak (it is a much larger effect for larger  dark matter halos 
with masses corresponding to those of rich groups and clusters).
Dark matter halos of galactic mass have typically accreted 15-20\% of their final
mass in the past 3 Gyr,  $\sim 10$\% in the past  2 Gyr and $\sim 5$\%
in the past  Gyr.
As well as the average accretion history, the simulations can be
used to study the {\em scatter} in the accretion histories among  different halos
of the same mass. The right panel of Figure 10 shows the r.m.s 
dispersion in the fraction of the final mass of the halo
that is contained in the largest progenitor as a function
of lookback time for halos in the same ranges of mass. 
As can be seen, the dispersion is also very insensitive to
the mass of the halo.  Figure 11 shows nine random accretion histories for
halos with masses between   $10^{12}$ and  $3\times 10^{12} M_{\odot}$.
As can be seen, there is a significant variation in recent accretion
histories. In some halos, there has been
very little accretion over the past several Gyr, while in others
the halo has accreted as much as a third of its mass during this
period. It can also be seen that in some haloes, the accretion occurs smoothly
over timescales of many Gyrs and in others, there are sudden ``jumps''
when the halo significantly increases its  mass over a short period of time.
\footnote {Note that numerical effects occasionally cause the masses
of halos to decrease by a small amount and a faithful representation would 
lead to a number of downward ``glitches'' being visible  in Figure 11. We
have chosen to smooth over these by forcing the mass to remain constant during
these periods. This allows us to transform directly from the
mass accretion history of the halo to the star formation history of the galaxy as
described in section 4.3} We intend to present a statistical analysis of these different
types of accretion events in a future paper.

\begin{figure}
\centerline{
\epsfxsize=14cm \epsfbox{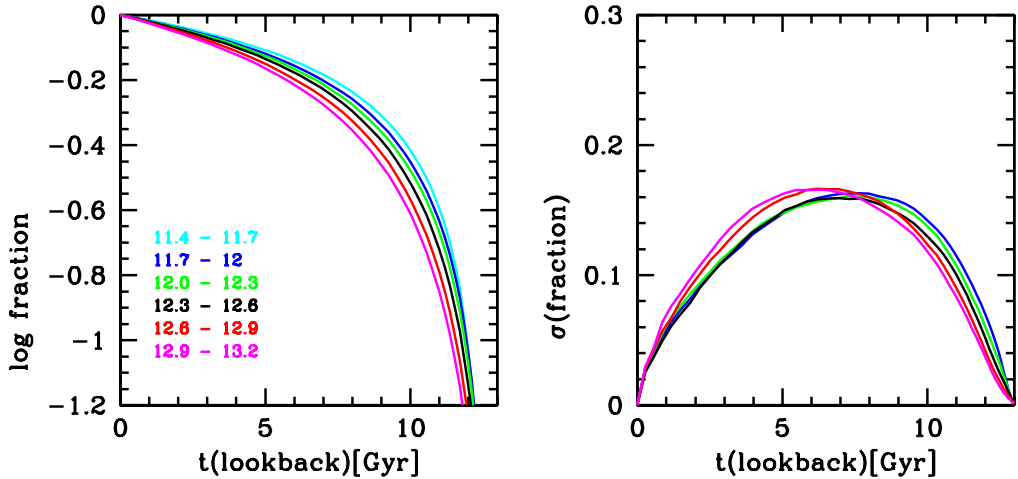}
}
\caption{\label{fig10}
\small
{\em Left:}The logarithm of the fraction of the final
mass of the halo contained in the main progenitor is plotted
as a function of lookback time  for dark matter  halos in 6 different
mass ranges as calculated from the {\em Millennium Run} simulation. {\em Right:}
The r.m.s. deviation in the fraction is plotted as a function of lookback
time. }
\end {figure}
\normalsize

\begin{figure}
\centerline{
\epsfxsize=9cm \epsfbox{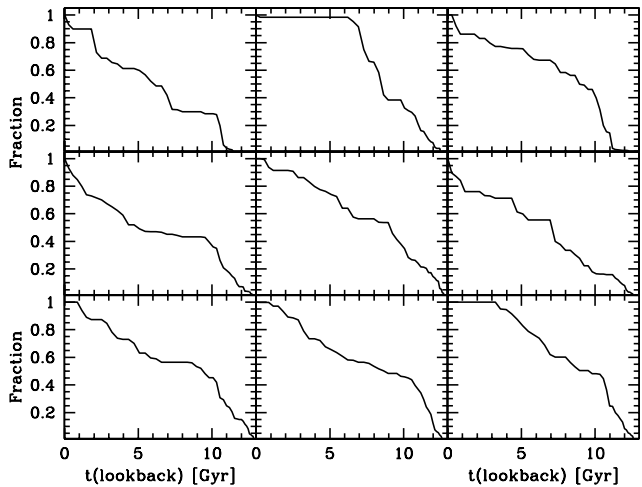}
}
\caption{\label{fig11}
\small
Examples of 9 different  mass accretion histories for dark matter  halos 
with masses in the range $10^{12}-3\times 10^{12} M_{\odot}$}
\end {figure}
\normalsize

\subsection {From mass accretion histories to star formation histories}

We now make the simple assumption that the star formation history of  a galaxy
directly tracks the accretion history of its surrounding  dark matter halo
(and hence the accretion history of cold gas).
The accretion histories are computed
directly from the Millennium Run simulation as described above.
The only free parameter in the model is the timescale over which
the accreted gas is converted into stars. We will denote this
gas consumption timescale as $t_{cons}$ from now on. During time
$\Delta t= t_{i+1}-t_i$, the halo will have accreted mass 
$\Delta M=M_{i+1}-M_i$, which corresponds to a fraction $f= \Delta M/M_{final}$ 
of the final mass of the halo. We assume the same fraction $f$ of
the final baryonic  mass of the galaxy is also accreted during this time step and that
it is transformed into stars at a constant rate over time $t_{cons}$.

As shown in Figure 10, the distribution of specific  accretion histories of dark matter 
halos varies very little with mass over the range 
$3 \times 10^{11} - 10^{13} M_{\odot}$. This means that $t_{cons}$
is the only parameter  that significantly affects the star formation histories
of galaxies in our model. At recent times, all dark matter halos accrete, on average, the
same  fraction   
of their mass. According to our simple model, all galaxies therefore accrete
the same fraction of their mass in gas (In practice, some fraction
of the accreted material will be in the form of stars, but we will
neglect this for simplicity).   
If the gas consumption time is short, the accreted gas  is converted 
into stars very rapidly in a burst. If the gas consumption time is
long, the effects of the individual accretion events will be smoothed out
and the star formation history of the galaxy will be more
continuous. This model is thus qualitatively
similar to scenario 2 described in section 3.2. The fact that the
average specific  accretion histories are so insensitive to the mass of the dark
matter halos provides 
a natural explanation for
why the  specific star formation rate averaged over a population of galaxies
should depend so weakly on mass (see discussion at the end of section 3.2).
We now investigate
whether the model can  yield results that are in {\em quantitative} agreement
with the observations.

Figure 12 shows the distribution of D$_n$(4000)  predicted by the models
for different values of the gas consumption timescale $t_{cons}$. 
These plots are for all halos 
with masses in the range $5 \times 10^{11} - 10^{12} M_{\odot}$, but
as discussed, very similar results would be obtained for other mass ranges.
We calculate these D$_n$(4000) values by summing the weighted contributions of a 
series of single age stellar populations (SSPs) to the red and the blue sides
of the index. The SSP predictions are generated using the GALEXEV software
(Bruzual \& Charlot 2003).

For galaxies with mean stellar ages greater than a few Gyr, D$_n$(4000)
is sensitive to metallicity as well as age. We have adopted a metallicity
of 0.5 solar, which is an appropriate mean metallicity for galaxies with stellar masses
$\sim 10^{10} M_{\odot}$ (Gallazzi et al 2005). As can be seen,
when $t_{cons}$ is large, the main body of the distribution
is peaked at D$_n$(4000)$\sim 1.3-1.4$. As $t_{cons}$ decreases,
the fraction of galaxies in the red tail increases, but the position
of the blue peak remains relatively constant. This
is very reminiscent of the trends seen in Figures 7 and 8 at 
stellar mass densities below $3 \times 10^8 M_{\odot}$ kpc$^{-2}$.
When $t_{cons}$ is small, the number of
galaxies in the red peak becomes almost equal to the number  
of galaxies in the blue peak. However, the 
D$_n$(4000) distribution  does not switch parity  and is never  peaked 
towards the red. In order to reproduce the D$_n$(4000)
distributions of galaxies with $\mu_* > 3 \times 10^8 M_{\odot}$ kpc$^{-2}$ shown
in Figures 7 and 8,  star formation must be decoupled from               
mass accretion at high stellar surface mass densities.

In Figure 13, we examine how the distribution of specific star formation
rates predicted by the model depends on the value of the
gas consumption time. The thick solid
line shows   the average star formation rate per unit stellar mass
as a function of $t_{cons}$  for our ensemble of  models. 
The upper curves show the 95th, 97th and 99th percentiles
of the distribution, 
while the lower curves show the lower 25th and 33rd percentiles of the distribution.
As can be seen, the trends as a function of decreasing gas consumption
time are very similar to the trends as a function of increasing stellar surface
mass density seen in Figure 9. The average
value of SFR/$M_*$ remains constant (the decrease
seen in the plot is because smaller fraction of the total
accreted gas has been converted into stars in models with
long gas consumption times).
The uppermost percentiles of the distribution increase by 0.1-0.2 dex
as $t_{cons}$ decreases  from values close to a  Hubble time to 0.1 Gyr. The lower
percentiles exhibit a strong drop in value, particularly for
$t_{cons} < 3$ Gyr. The average specific star formation rate is slightly  (0.1-0.2 dex)
lower than in the observations. This may indicate that baryons were converted
into stars less efficiently at high redshifts than at present.

\begin{figure}
\centerline{
\epsfxsize=15cm \epsfbox{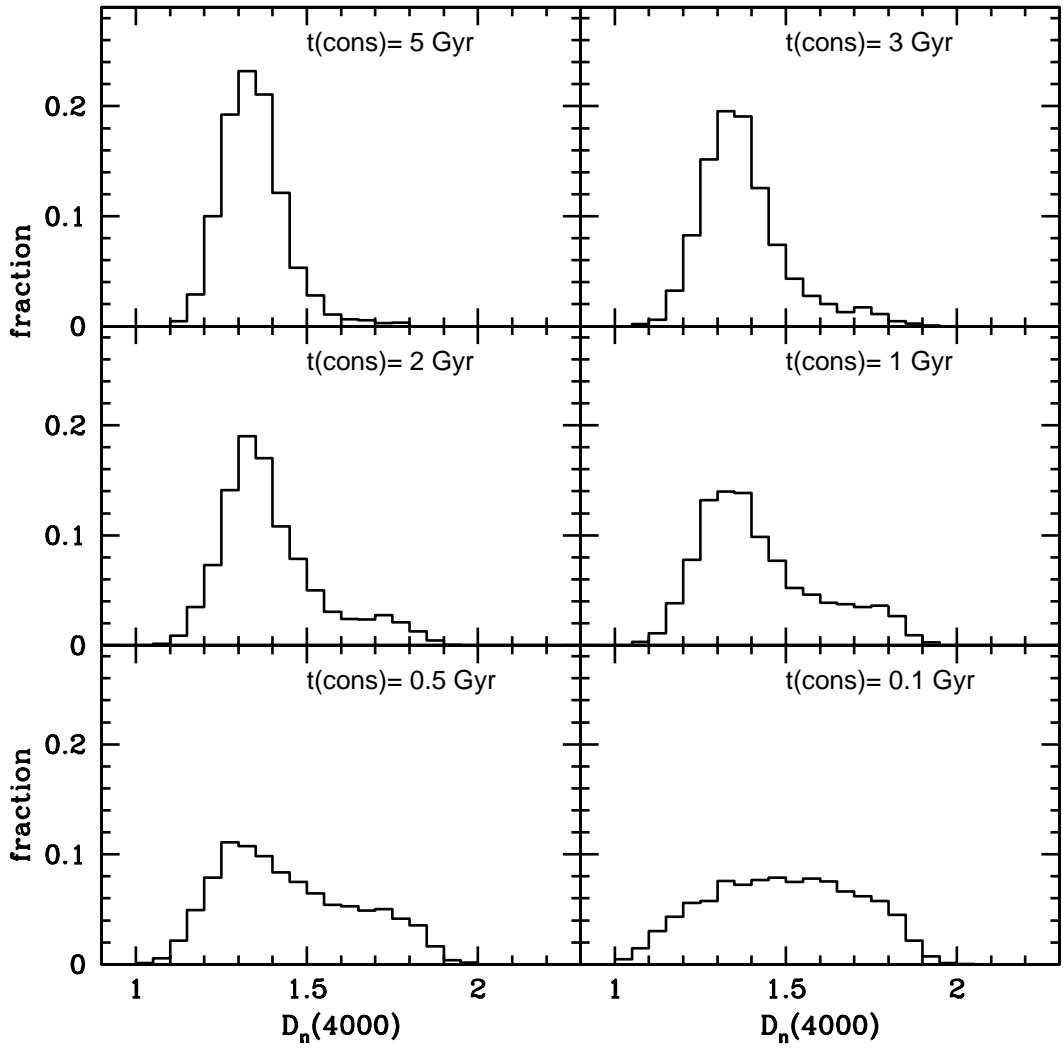}
}
\caption{\label{fig12}
\small
The distribution of D$_n$(4000) predicted by the model is plotted for
6 different choices of gas consumption time $t_{cons}$.}
\end {figure}
\normalsize

\begin{figure}
\centerline{
\epsfxsize=7cm \epsfbox{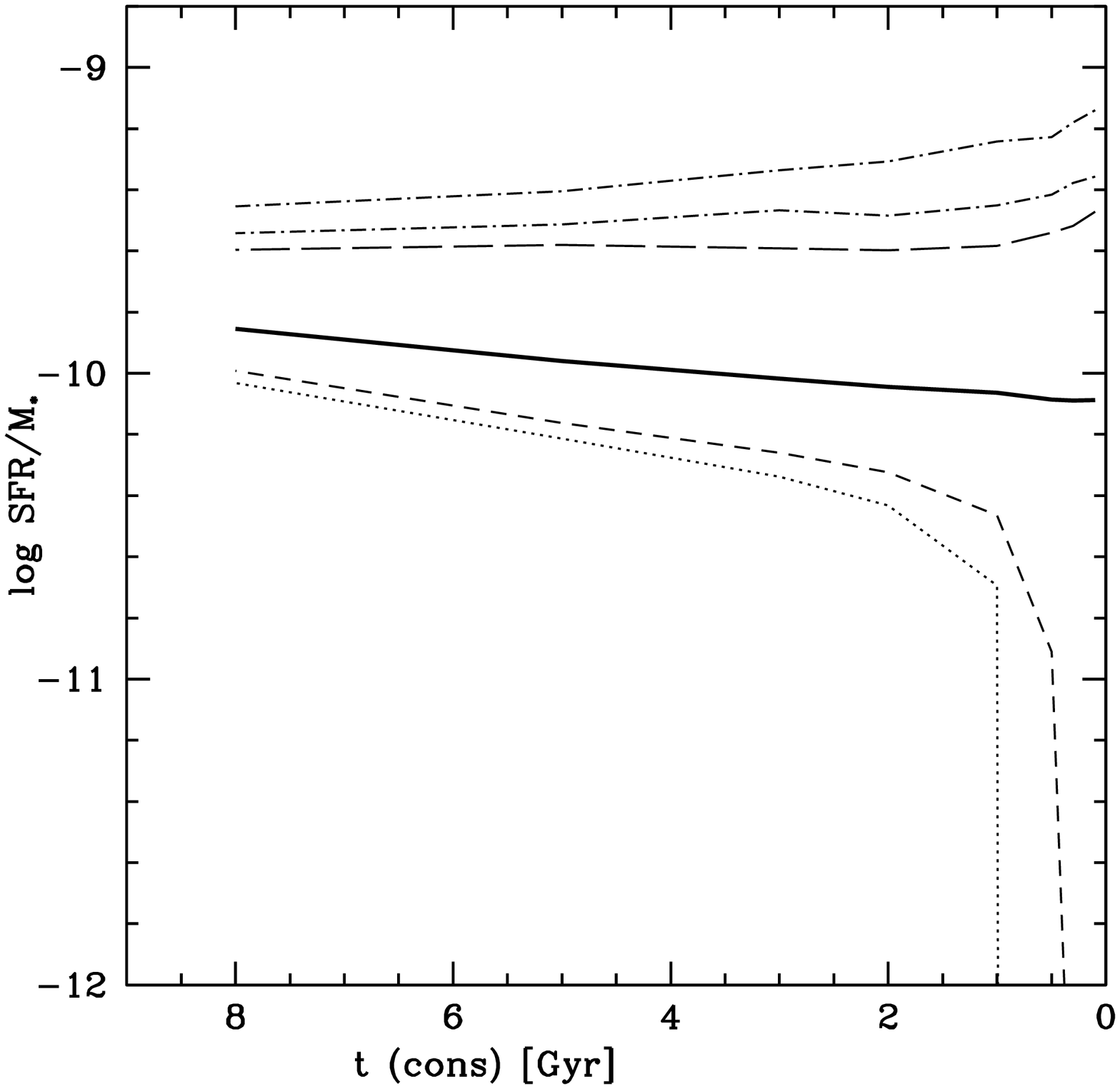}
}
\caption{\label{fig13}
\small
The average value of $\log$ SFR$/M_*$ ) predicted by our model
based on dark halo accretion histories is plotted as a function
of gas consumption time $t_{cons}$ (thick solid line). The upper curves show the
95th, 97th and 99th percentiles of the distribution of $\log$ SFR/$M_*$
as a function of $t_{cons}$, while the lower curves show the
25th and 33rd percentiles of this distribution. }
\end {figure}
\normalsize

We now turn the analysis around and ask whether we can use the models
to constrain how the inferred gas consumption times scale  with the observed 
stellar surface mass densities of the galaxies in our sample.
Our results are
shown in Figure 14.  The left hand panels show the variation in
D$_n$(4000) and H$\delta_A$ for galaxy pairs              
predicted by the models for different values
of $t_{cons}$. In the right hand panels,
we plot the variation as a function of $\mu_*$ for the SDSS
galaxies. Red symbols correspond to galaxies with $\log M_*$ in the
range 10-10.25, green symbols to 10.25-10.5 and black symbols
to 10.5-10.75. As can be seen, galaxies with stellar surface densities
of $\sim 10^8 M_{\odot}$ kpc$^{-2}$ have inferred gas consumption
times of around 4-5 Gyr. At stellar surface densities close to
the critical value of $3 \times 10^8 M_{\odot}$ kpc$^{-2}$, the
inferred timescales have decreased to less than a Gyr.

To see if this behaviour can be easily understood, we 
make the following assumptions based on the approximation that the
variation in properties is driven by individual, well-separated
accretion events whose typical size scales with
the mass of the galaxy (see Figure 10):
\begin {enumerate}
\item The mean surface density of the accreted gas scales in direct
proportion to the mean surface density of the stars that are
already present in the galaxy.
\item There is a power-law dependence between the mean surface
density of star formation and the mean surface density of the accreted gas 
(i.e. the star formation obeys a global Schmidt law).
\item All of the gas is consumed into stars; none of it is ejected.
\end {enumerate}
The Schmidt law (equation 1) then allows one to derive a relation between
the gas consumption time $t_{cons}$  
and $\Sigma_{gas}$
($t_{cons}= M_{gas}/\dot{M}_{gas} \propto
\Sigma_{gas}/ \dot{\Sigma}_{\rm SFR} \propto \Sigma_{gas}^{1-N}$).
By assumption 1, we have that   $ \Sigma_{gas} \propto \mu_*$,
so that $t_{cons} \propto \mu_*^{1-N}$.
The curves shown in the left panels of Figure 14 relate the observed variation
in D$_n$(4000) and H$\delta_A$ to the gas consumption time.
The two curves on the right hand panels show the scalings between
the variation in D$_n$(4000)/H$\delta_A$ and $\mu_*$  
that are then predicted for different values of the Schmidt-law exponent $N$.
The dotted curve
is for a Schmidt law with $N=1.5$, while the long-dashed
curve shows the relation  expected if $N=2$.
The curves are  normalized to go through the same point
at $\log \mu_* =8$.
As can be seen, the observations are better fit by a Schmidt law model
with $N=2$, where the consumption time   
scales as $t_{cons} \propto \Sigma_{gas}^{-1} \propto  \mu_*^{-1}$. 
We caution, however, that our methodology only provides 
indirect constraints on the form of the star formation law.
All three assumptions listed above are subject to considerable
uncertainty. 

In models for the formation of galactic discs by the
condensation of gas in gravitationally dominant dark matter 
halos (Fall \& Efstathiou 1980; Mo, Mao, White 1998)
the density of the accreted gas is determined
by the angular momentum of the surrounding  halo.
Analysis of the size distributions of galaxies in the SDSS (Kauffmann et al 2003b;
Shen et al 2003) shows that at a given luminosity or stellar mass, the distribution
of half-light radii is well described by a log-normal function.
For low mass galaxies, the width of the log-normal function is in good
agreement with theoretical expectations. This suggests that the
disk formation models do provide a reasonable explanation for the
observed sizes of galaxies. So long as the angular momentum 
distributions of the dark matter halos surrounding galaxies of fixed mass and size 
have not changed substantially over the past few Gyr, one might
expect assumption (1)  to be valid.

Assumption (2) is  more problematic.
The Schmidt law is known to break down
when the surface density of gas falls low enough so that the disk is             
stabilized against the effects of 
gravitational perturbations (Van der Hulst et al 1993). 
The effective gas consumption
times in low surface density galaxies are thus likely to be
longer than predicted by the canonical
Schmidt law with $N=1.5$.
In addition, the  Schmidt Law only comes into play once the
accreted gas has settled into a cold disk and the  timescale for this to
happen may not be  the same as the timescale over which star formation occurs
once the disk is present. Finally, assumption (3) breaks down if a
substantial fraction of the accreted gas is expelled from
the galaxy by supernovae-driven winds.

In spite of these ambiguities, we find it encouraging that our 
simple model is able to reproduce so many of the trends seen in the
observations self-consistently and with a minimum of free parameters. 

\begin{figure}
\centerline{
\epsfxsize=15cm \epsfbox{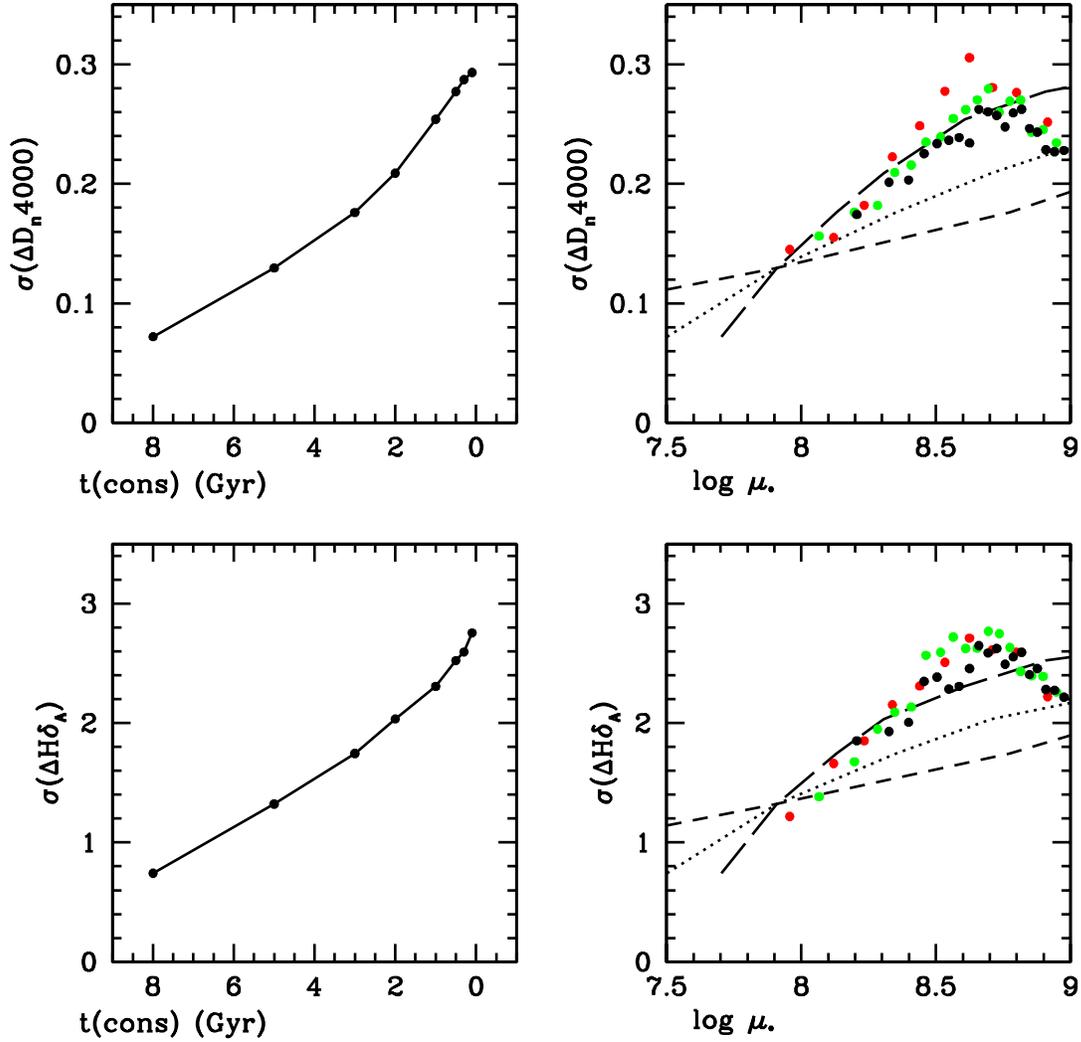}
}
\caption{\label{fig14}
\small
{\bf Left:} The r.m.s. 
difference between D$_n$(4000) and H$\delta_A$ for galaxy pairs
as a function of $t_{cons}$ in the models. {\bf Right:} The r.m.s. difference                
as a function of $\mu_*$ for  SDSS
galaxies. Red symbols correspond to galaxies with $\log M_*$ in the
range 10-10.25, green symbols to 10.25-10.5 and black symbols
to 10.5-10.75. The dotted and long-dashed curves show
the relations expected for Schmidt laws with N=1.5 and 2 (see text).} 
\end {figure}
\normalsize

\section {Summary and Discussion}

In this paper, we have attempted to infer the nature of the recent                 
star formation histories of local galaxies by studying the scatter
in their spectral properties. We have presented evidence that 
the distribution of star formation histories changes qualitatively above a             
characteristic stellar 
surface mass density of $3 \times 10^8 M_{\odot}$ kpc$^{-2}$.
Below this  density, the mean star formation rate of galaxies of given
mass and surface density is independent of the values of mass and surface
density, while the scatter in star formation rate {\em increases} with
surface density. The simplest interpretation is that all galaxies
in this regime are forming stars at a rate which will typically double
their mass in a Hubble time, but that the star formation
in the denser galaxies is more bursty.
As a result,  the scatter in specific star formation rates,           
stellar absorption line indices and colours  is larger 
for smaller, more concentrated galaxies.                     
Above the characteristic surface density, galaxy growth through star formation
shuts down  and the scatter in 
spectral properties decreases.

We propose that in galaxies below the characteristic stellar
surface mass density of $3 \times 10^8 M_{\odot}$ kpc$^{-2}$,
star formation events  are triggered               
when cold  gas is accreted onto a galaxy. 
We have used a new high resolution  numerical
simulation of structure formation in a ``concordance'' $\Lambda$CDM
Universe to quantify the incidence of these accretion events                
showing that observational data are  well fit by a model
in which the consumption time of accreted  gas decreases with the surface
density of the galaxy  as $ t_{cons} \propto \mu_*^{-1}$.
In high density galaxies, star formation ceases to be coupled to 
the hierarchical build-up of dark matter halos.

One question that arises is whether these results and inferences are consistent with
past studies of the star formation histories of nearby galaxies.
Qualitatively, at least, we believe that the answer is yes. 
One area where the issue of the intrinsic scatter in the luminosities and colours
of galaxies has been of considerable importance is in the establishment
of the extragalactic distance scale.
The use of galaxies as distance indicators is dependent on the accuracy
with which one can predict the intrinsic luminosity of the system from other
measurable quantities such as its  rotation speed or stellar velocity
dispersion. In order for  galaxies to be good distance indicators,
the intrinsic scatter in their stellar mass-to-light ratios must be small.
Historically, the use of galaxies as distance indicators has
been confined to two classes of object: ellipticals,
which have high stellar surface densities, 
and late-type (i.e. low density ) spirals. As shown in this paper,
this is where the variation in spectral properties is at a minimum.   
What has not been extensively discussed in the literature           
is the fact that the variation 
exhibits a peak in the regime of parameter space between the 
two galaxy classes.                                  

It is also interesting to consider whether the conclusions in this 
paper can shed light on trends in  the star formation histories
among different kinds of dwarf galaxies.
Van Zee (2001) used
UBV photometry,  H$\alpha$ imaging and  HI observations
of a large sample of isolated dwarf irregular galaxies
to argue that these systems have experienced slow, but constant star
formation over  timescales comparable to a Hubble time.
The derived gas depletion timescales were also very long,
typically $\sim 20$ Gyr. These star formation histories are in strong contrast
to those of blue compact
dwarf galaxies, which have long been known to be  
experiencing strong bursts of star formation.

If blue compact galaxies are bursting because they have recently experienced a gas
infall event, there should be 
evidence that the HI gas distribution and kinematics are
systematically different in these systems  than in ordinary
dwarf irregulars. The HI-to-optical diameters of
BCGs are typically larger than those of normal
irregular galaxies (Gordon \& Gottesman 1981).
 Whereas most dwarf irregular galaxies have
well-defined, low-dispersion rotating disks of HI gas
that appear undisturbed ( Skillman et al 1987; van Zee et al 1997), the
HI kinematics of blue compact galaxies is much more complex. Some
BCGs have extensive HI haloes that appear to be rotating (Brinks \& Klein 1988;Meurer et al 1996).
The 5 BCGs studied by van Zee, Skillman \& Salzer (1998) all have non-ordered
kinematic structure.  Gordon \& Gottesman (1981) and Stil \& Israel (2002) 
argue that the HI gas
in the extended envelopes around these galaxies may be falling in and
providing the fuel for the starburst.  If the infalling
material has not yet been enriched, it would help explain why the 
metallicities of BCGs are frequently lower than those of
other dwarf galaxies of the same luminosity (see for example Werk,
Jangren \& Salzer 2004).
I Zw 18  is perhaps the most famous example of a blue compact
galaxy with infalling gas -- it has an HI envelope that extends
a factor of 8 beyond its optical radius (Lequeux \& Viallefond 1980) 
and a metallicity only
1/38th of solar (Searle \& Sargent 1972; Skillman \& Kennicutt 1993). 
Although these results for dwarf galaxies are tantalizing and       
suggest that our infall model may provide a way of understanding
the star formation cycles in these systems, the lack of large, complete
samples of dwarfs  makes it difficult reach  any  definitive conclusions.  

Further checks on our  model will come from the analysis of complete
samples of galaxies with HI  gas mass  measurements 
as well as star formation rates and absorption line indices
from optical spectra. If our model is correct, then the trends in the
gas mass fractions of galaxies as a function of stellar mass and stellar
surface mass density ought to mirror the trends in             
specific star formation rates shown in Figure 9.                     
As surface density increases, we expect galaxies of fixed stellar mass to
use up the infalling gas more rapidly and so to have correspondingly
lower mean gas fractions.

It is also interesting to investigate how galaxies would be expected to
evolve as a function of redshift using our infall models. As seen
in the left panel of Figure 10, the infall rates
are predicted to  increase strongly as a function of
lookback time. As discussed in section 4.2, local galaxies have 
only accreted 5\% of their mass
in the past Gyr on average. At redshift 1, an average of 20\% of the mass was accreted
in the last Gyr and by redshift 2, nearly half the mass of a typical galaxy
was assembled over this same period. Rather than the present-day trickle, the infall of 
gas would have resembled  a flood. We intend to explore the implications
of this in more detail in future work.

Finally, we have not addressed the reason why star formation switches off
in galaxies with  stellar surface mass densities greater than
$3 \times 10^8 M_{\odot}$ kpc$^{-2}$. As shown in Figure 10, mergers and accretion 
are important in dark matter halos of all masses -- indeed, more massive halos
are expected to have accreted a larger fraction of their  mass at late times.
The observations indicate, however, that star formation in massive and dense galaxies
cannot be linked to these events any longer. The results shown in Figure 15
may provide a clue to this conundrum. We have plotted the average
specific star formation rate $\overline{SFR}/M_*$ as a function of
stellar surface density $\mu_*$ for galaxies with stellar masses
in the range $ 3 \times 10^9 - 3 \times 10^{10} M_{\odot}$.
The galaxies have been split into five different concentration bins. 
At stellar surface densities less than $3 \times 10^8 M_{\odot}$, 
$\overline{SFR}/M_*$ increases for more concentrated galaxies.
This is presumably because the starburst is often concentrated towards
the centre of the galaxy.
At stellar surface densities greater than $3 \times 10^8 M_{\odot}$, 
$\overline{SFR}/M_*$  declines  for more concentrated
galaxies. The obvious interpretation of this result is that 
star formation is increasingly suppressed as the bulge of the galaxy becomes more
dominant. 

\begin{figure}
\centerline{
\epsfxsize=9cm \epsfbox{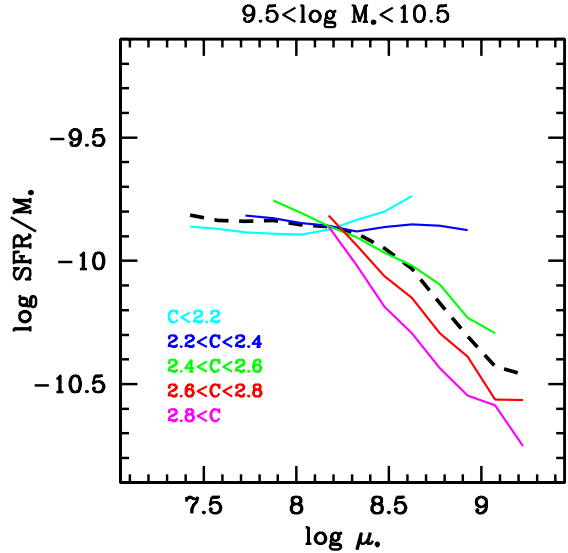}
}
\caption{\label{fig15}
\small
The average star formation rate divided by
the stellar mass within the fibre aperture is plotted against the logarithm of
the stellar surface density for galaxies in the mass range 
$3\times 10^9 - 3 \times 10^{10} M_{\odot}$. The thick dashed line shows
the result for all galaxies, while the coloured lines show results
for galaxies in differenr ranges of concentration.}
\end {figure}
\normalsize

One possibility is that gas in the halo is
heated and possibly even expelled as the bulge is formed 
(Granato et al 2001; Springel, Di Matteo, Hernquist 2005).
In this case, bulge-dominated galaxies would continue
to merge, but with very little associated star formation. This
has been dubbed ``dry merging'' in recent papers by Faber et al (2005)
and Van Dokkum et al (2005). It is also interesting the emission from 
active galactic nuclei (AGN) in the local Universe peaks very close to
the stellar surface mass density ($3 \times 10^8 M_{\odot}$ kpc$^{-2}$)
where the switch in  star formation behaviour occurs (Heckman et al 2004).  
This coincidence raises the possibility that the two phenomena are
causally connected, although the  empirical demonstration that this is the case
still remains as a challenge for the future.

\vspace{1.2cm}
\large
\noindent
{\bf Acknowledgements}\\
\normalsize

\noindent
G.~D.~L. thanks the Alexander von Humboldt Foundation, the Federal Ministry of
Education and Research, and the Programme for Investment in the Future (ZIP) of
the German Government for financial support.

Funding for the creation and distribution of the SDSS Archive has been 
provided by the Alfred P. Sloan Foundation, the Participating Institutions, 
the National Aeronautics and Space Administration, 
the National Science Foundation, the U.S. Department of Energy, 
the Japanese Monbukagakusho, and the Max Planck Society. 
The SDSS Web site is http://www.sdss.org/.
The SDSS is managed by the Astrophysical Research Consortium (ARC) 
for the Participating Institutions. The Participating Institutions 
are The University of Chicago, Fermilab, the Institute for Advanced Study, 
the Japan Participation Group, The Johns Hopkins University, 
the Korean Scientist Group, Los Alamos National Laboratory, 
the Max-Planck-Institute for Astronomy (MPIA), 
the Max-Planck-Institute for Astrophysics (MPA), 
New Mexico State University, University of Pittsburgh, 
University of Portsmouth, Princeton University, 
the United States Naval Observatory, and the University of Washington.

\pagebreak
\Large
\begin {center} {\bf References} \\
\end {center}
\normalsize
\parindent -7mm
\parskip 3mm

Adelman-McCarthy, J. et al., 2004, ApJS, in press      

Arp, H., 1966, ApJS, 14, 1 

Birnboim, Y., Dekel, A., 2003, MNRAS, 345, 349

Balogh, M.L., Morris, S.L., Yee, H.K.C., Carlberg, R.G., Ellingson, E., 1999,
ApJ, 527, 54

Blanton, M.R., Lupton, R.H., Maley, F.M., Young, N., Zehavi, I., \&  
   Loveday, J. 2003a, AJ, 125, 2276 

Blanton, M.R., et al., 2003b, ApJ, 594, 186

Blanton, M.R., Eisenstein, D., Hogg, D.W., Schlegel, D.J., Brinkmann, J., 2005, ApJ, 629,143

Brinchmann, J., Charlot, S., White, S.D.M., Tremonti, C., Kauffmann, G.,
Heckman, T., Brinkmann, J., 2004, MNRAS, 351, 1151

Bruzual, G., Charlot, S., 2003, MNRAS, 344, 1000

Charlot, S., Silk, J., 1994, ApJ, 432, 453

Cole, S., Aragon-Salamanca, A., Frenk, C.S., Navarro, J.F. Zepf, S., 1994, MNRAS, 271, 781

Cole, S., Lacey, C.G., Baugh, C.M., Frenk, C.S., 2000, MNRAS, 319, 168

Croton, D., 2005, MNRAS, submitted (astro-ph/0508046)

De la Fuente Marcos, R., De la Fuente Marcos, C., 2004, NewA, 9, 475

Dolphin, A.E., 2000, MNRAS, 313, 281

Dolphin, A.E., Weisz, D.R., Skillman, E.D., Holtzman, J.A., 2005,
astro-ph/0506430 

Faber, S.M. et al, 2005, ApJ, submitted (astro-ph/0506044)

Fall, S.M., Efstathiou, G., 1980, MNRAS, 193, 189

Forcada-Miro, M.I. \& White, S.D.M., 1997, astro-ph/9712204

Fukugita, M., Ichikawa, T., Gunn, J.~E., Doi, M., Shimasaku, K., 
\& Schneider, D.~P., 1996, AJ, 111, 1748 

Gallazzi, A., Charlot, S., Brinchmann, J., White, S.D.M., 
Tremonti, C.A., 2005, MNRAS, 362, 41

Gerola, H., Seiden, P.E., Schulman, L.S., 1980, ApJ, 242, 517

Gordon, D., Gottesman, S.T., 1981, AJ, 86, 161

Granato, G.L., De Zotti, G., Silva, L., Bressan, A., Danese, L., 2004, ApJ, 600, 580

Grebel, E.K., 2001, in "Dwarf galaxies and their environment", 
Shaker Verlag,  Eds. De Boer,K.,Dettmar,R.,Klein,U.. , p.45

Gunn, J.~E.~et al., 1998, AJ, 116, 3040 

Gunn, J.E. et al. 2005, AJ, submitted

Harris, J., Zaritsky, D., 2004, AJ, 127, 1531

Heckman, T.M., Kauffmann, G., Brinchmann, J., Charlot, S., Tremonti, C., 
White, S.D.M., 2003, ApJ, 613, 109

Hernandez, X., Valls-Gabaud, D. Gilmore, G., 2000, MNRAS, 316, 605

Hogg, D.~W., Finkbeiner, D.~P., Schlegel, D.~J., and Gunn, J.~E., 2001,
AJ, 122, 2129 

Hogg, D.W. etal al, 2003, ApJ, 585, L5

Ivezic, Z.R et al, 2004, AN, 325, 583

Kauffmann, G., White, S.D.M., Guiderdoni, B., 1993, MNRAS, 264, 201

Kauffmann, G., Colberg, J., Diaferio, A., White, S.D.M., 1999, MNRAS, 307, 529

Kauffmann, G, et al, 2003a, MNRAS, 341, 33

Kauffmann, G, et al, 2003b, MNRAS, 341, 54

Kauffmann, G., White, S.D.M., Heckman, T.M., Menard, B., Brinchmann, J., Charlot, S., 
Tremonti,C., Brinkmann, J., 2004, MNRAS, 353, 713

Kennicutt, R.C., 1998, ApJ, 498, 181

Lequeux, J., Viallefond, F., 1980, A\&A, 91, 261

Mo, H.J., Mao, S., White, S.D.M., 1998, MNRAS, 295, 319

Percival, W.J. et al, 2001, MNRAS, 327, 1297

Pier, J.R., Munn, J.A., Hindsley, R.B., Hennessy, G.S., Kent, S.M.,
Lupton, R.H., and Ivezic, Z. 2003, AJ, 125, 1559 

Sandage, A., The Hubble Atlas of Galaxies, Washington: Carnegie Institution, 1961

Schlegel, D.~J., Finkbeiner, D.~P., \& Davis, M., 1998, ApJ, 500, 525

Schmidt, M., 1959, ApJ, 129, 243

Searle, L., Sargent, W., 1972, ApJ, 173, 25

Searle,L, Sargent, W., Bagnuolo, W.G., 1973, ApJ, 179, 427

Seiden, P.E., Gerola, H., 1979, ApJ, 233

Shen, S. et al, 2003, MNRAS, 343, 978

Shimasaku, K.~et al., 2001, AJ , 122, 1238 

Skillman, E.D., Bothun, G.D., Murray, M.A., Warmels, R.H., 1987, A\&A, 185, 61

Skillman, E.D., Kennicutt, R.C., 1993, ApJ, 411, 655

Smecker-Hane,T.A., Cole, A.A., Gallagher, J.S., Stetson, P.B., 2002, ApJ, 566, 239

Smith, J.~A.~et al., 2002, AJ, 123, 2121 

Somerville, R.S., Primack, J., 1999, MNRAS, 210, 1087

Spergel, D. et al, 2003, ApJS, 148, 175

Springel, V., White, S.D.M., Tormen, G., Kauffmann, G., 2001, MNRAS, 328, 726

Springel, V. et al., 2005, Nature, 435, 629

Springel, V., Di Matteo, T., Hernquist, L., 2005, MNRAS, 361, 776

Stil, J.M., Israel, F.P., 2002, A\&A, 392, 473

Stoughton, C.~et al., 2002, AJ, 123, 485 

Strateva, I. et al 2001, AJ, 122, 1861

Strauss, M.~A.~et al., 2002, AJ, 124, 1810 

Tegmark, M. et al, 2004, ApJ, 606, 702

Tinsley, B.,  1968, ApJ, 151, 547

Tremonti, C.A. et al, 2004, ApJ, 613, 898

Tucker, D et al, 2005, AJ, submitted

Van der Hulst, J.M., Skillman, E.D., Smith, T.R., Bothun, G.D., McGaugh, S.S., 
De Blok, W.J.G., 1993, AJ, 106, 548

van Dokkum, P., 2005, AJ, in press (astro-ph/0506661)

van Zee, L., Maddalena, R.J., Haynes, M.P., Hogg, D.E., Roberts, M.S., 1997, AJ, 113, 1638

van Zee, L., 2001, AJ, 121, 2003

van Zee, L., Skillman, E.D., Salzer, J.J., 1998, AJ, 111,1186

Werk, J.K., Jangren, A., Salzer, J.J., 2004, ApJ, 617, 1004

White, S.D.M., Rees, M.J., 1978, MNRAS, 183, 341

White, S.D.M., Frenk, C.S., 1991, ApJ, 379, 52     

Worthey, G., Ottaviani, D.L., 1997, ApJS, 111, 377

York, D.~G.~et al., 2000, AJ, 120, 1579

\end{document}